# Recent Advances in Novel Materials and Techniques for Developing Transparent Wound Dressings†

Muzammil Kuddushi,[a] Aatif Ali Shah,[a] Cagri Ayranci,[b] and Xuehua Zhang[a]

**Abstract** Optically transparent wound dressings offer a range of potential applications in the biomedical field, as they allow for the monitoring of wound healing progress without having to replace the dressing. These dressings must be impermeable to water and bacteria, yet permeable to moisture vapor and atmospheric gases in order to maintain a moist environment at the wound site. This review article provides a comprehensive overview of the types of wound dressings, novel wound dressing materials, advanced fabrication techniques for transparent wound dressing materials, and the key features and applications of transparent dressings for the healing process, as well as how it can improve healing outcomes. This review mainly focuses on representing specifications of transparent polymeric wound dressing materials, such as transparent electrospun nanofibers, transparent crosslinked hydrogels, and transparent composite films/membranes. Due to the advance properties of electrospun nanofiber such as large sur-face area, enable efficient incorporation of antibacterial molecules, a structure similar to the extracellular matrix, and high mechanical stability, is often used in wound dressing applications. We also highlight the hydrogels or films for wound healing applications, it's promote the healing process, provide a moisture environment, and offer pain relief with their cool, high-water content, excellent biocompatibility, and bio-biodegradability. But as the hydrogels or films fabricated with a single component have low mechanical strength and stability, recent trends have offered composite or hy-brid materials to achieve the typical wound dressing requirements. Advanced wound dressings with transparency, high mechanical stability, and antimicrobial functionality are becoming a popular research avenue in the wound dressing research field. Finally, the developmental prospectives of the new transparent wound dressing materials for future researches are presented.

## 1 Introduction

Skin is a crucial organ of the human body and serves as a key barrier against external infections. Every year, numerous people suffer from various types of skin injuries or burns caused by fires, accidents, boiling oil, and water [1,2]. Proper wound care is critical, regardless of whether the wound is a small cut or a large incision, as poor wound healing may result in infection, various patholog-ical reactions, or even death [3–6]. When selecting a dressing mate-rial for treatment, factors such as wound size, type, location, level of bacterial infection, amount of exudate, and adhesive proper-ties should be considered, and the dressing material should have compatible characteristics with the skin to promote the healing process [7].

An ideal wound dressing should be biocompatible, protect the injured site, absorb exudate, accelerate healing activities, keep the wound moist, and allow gases and fluids to exchange [8]. Sim-ilar to the extracellular matrix (ECM), a suitable wound dressing can provide a desirable environment for wound healing [9,10]. An ideal wound dressing should protect the wound from the exter-nal environment, exchange air, absorb and remove a high volume of exudate, have antibacterial properties against microorganisms and infections, decrease dead cells of a wound, provide strong mechanical protection, be easy to replace, be sterile, non-toxic and non-allergic, and relieve pain from an infected wound [6,11]. Modern wound dressings are fabricated using natural, synthetic, and biopolymer-based materials [12].

In the past decade, transparent nanofiber membranes have gained much attention due to their simple operation method, flex-ible design, controllable structure, unique excellent optical prop-erties, high specific surface area, and porosity [13]. Nanofibers con-taining different active compounds, such as antimicrobial agents, can help in wound healing, prevent dehydration, and infection. However, designing a multifunctional wound dressing with ideal features, including antibacterial, mechanical strength, high wa-ter vapor transmission rate, moist wound environment, good swelling ratio, transparency, and good permeability, still poses many challenges. The key objective of this review was to up-date and summarize current trends in applications and advances in transparent wound dressing materials, by means of the most used materials, such as crosslinked hydrogels, composite films or membranes, and electrospun nanofibers. With reference to the Web of Science database, the transparent dressing materials re-search literature from 2019 to 2022 is reviewed both quantita-tively and qualitatively.

[a] *Department of Chemical and Materials Engineering, University of Alberta, Alberta T6G 1H9, Canada; E-mail: Xuehua.Zhang@ualberta.ca*
[b] *Department of Mechanical Engineering, University of Alberta, Edmonton AB T6G 2H5, Canada.*

Transparent wound dressings have been prepared by various fabrication techniques. The transparency of wound dressings allows the visualization of the wound bed and effectively monitors the wound site, minimizing the patient's pain and risk of skin in-jury [14]. Since 2022, several review articles have been published on different kinds of wound dressings [15], such as hydrogels [16,17], electrospun scaffolds [18], alginate-based [19], and composite dressings [20], and the factors that influence the wound healing process. In addition, various bioengineered and synthetic approaches have been discussed in recent literature that protect wounds against microorganisms, promote tissue repair, and improve the healing process [4].

Dong et al. reviewed smart wound dressings as an alternative to conventional wound dressings that can interact with wounds and respond to the wound condition by employing in-built de-vices and/or smart materials, such as stimuli-responsive materi-als with self-healing characteristics [21]. In another work, Zhang et al. assessed hydrogel-based wound dressings with a focus on their adhesion feature and their classification depending on ad-hesive mechanisms [22]. Similarly, Gao et al. provided a systematic review of electrospun nanofiber-based wound dressings and how efficiently the electrospinning technology has been employed to develop nanofiber scaffolds with great healing properties due to the tailorability of the electrospinning process [23].

In this comprehensive review, we will discuss the recent advances and challenges in the field of transparent wound dressings. We will focus on the design and production of novel dressings that promote the healing process, and will classify the various materials used for their construction, including fibers, gels, and films. Additionally, we will examine the characteristics that contribute to the ideal transparent wound dressing, including bio-activity, biocompatibility, transparency, mechanical stability, and mainte-nance of a moist environment [24,25]. We will also highlight the promising effectiveness of transparent wound dressings in recent years and the progress made in their development. By the end of this review, readers will have a better understanding of the latest research in the field of wound restoration and the potential for novel transparent wound dressings to improve patient outcomes.

## 2 Classification of Wound Dressings

Chronic wounds are becoming increasingly prevalent and diffi-cult to treat. Therefore, it is crucial to use appropriate wound dressing materials such as synthetic polymers, elastomers, and natural polymers [26–28]. Wound dressings are categorized into six types: passive/traditional, interactive/modern, skin substitutes, bioactive, composite, and medicated wound dressings [29–31].

### 2.1 Passive /Traditional Wound Dressings

Passive/traditional dressings serve as a protective layer to restore the function of the skin underneath, such as gauze, plaster, and wool dressings. However, traditional dressings may cause infec-tion and slow down the healing process by transferring moisture molecules and gases through the dressings [32,33]. They are recom-mended for dry wounds since they stick properly to the wound but may cause severe pain during detachment. In contrast, mod-ern wound dressings have more advanced formulations that not only cover the wound but also prevent dehydration and promote the wound healing process [34].

### 2.2 Interactive/Modern Wound Dressings

Interactive dressings create a moist environment around the wound and stimulate the healing process by being permeable to water vapor and oxygen molecules, but not to the bacteria. In-teractive dressings include semi-permeable film dressings, foams, semi-permeable foam dressings, hydrogel dressings, and hydro-colloid dressings [35]. Choosing the appropriate modern dressing product can be challenging, given the numerous products avail-able in the market. The advantages of Interactive/Modern wound dressings are that they are inexpensive and reliable dressings with a longer shelf life and suitable for less exuding wounds [36,37].

### 2.3 Skin Substitutes Wound Dressings

Skin substitutes are made up of epidermal and dermal layers de-veloped from keratinocytes and fibroblasts on a collagen matrix, such as autografts, acellular xenografts, and allografts. However, they are not commonly used in wound healing applications due to some limitations, such as host rejection, short shelf life in the wound environment, and potential for infection transmission [38].

### 2.4 Bioactive Wound Dressings

Bioactive wound dressings are prepared from biopolymers and aimed at delivering encapsulated active substances (antibiotics, peptides, drugs, vitamins, growth factors etc.) to the wound en-vironment to enhance the process of wound healing. These active substance dressings improve wound healing activity by making an active interaction between the dressing and wound environment, whereas interactive or modern wound dressing materials directly interact with the wound bed promoting the regeneration process, and these interactions include removal of excessive exudate, pro-viding a moist environment in the wound bed, and prevention of infections. [39]. Natural materials, including hydrocolloids, al-ginates, collagens, chitosan, chitin, derivatives from chitosan or chitin, bio textiles, are commonly used in bioactive dressings, given their biocompatibility, biodegradability, and non-toxic na-ture.

Several studies have developed bioactive wound dressings us-ing natural materials. For example, Singla et al. developed nano bio-composites containing plants, showing that biomaterial-based wound dressings are ideal for rapid skin repair, decreasing the production of inflammatory cytokines and tissue regeneration [40]. Tang et al. reported honey/alginate/PVA nanofibrous membrane by electrospinning technique, which effectively constrained bac-

2 | 1–23

terial growth and showed an enhanced antibacterial effect [41][40]. Ionescu et al. developed a chitosan-based nanofiber, which exhibited considerable antioxidant and antimicrobial characteristics, making it appropriate for the treatment of chronic wounds [42].

### 2.5 Composite Wound Dressings

Composite wound dressings consist of multiple layers, each with unique biological properties. These dressings typically include three layers and are designed to adhere to skin tissues using either a transparent film or a non-woven fabric tape border. Composite dressings can be used as primary or secondary dressings and are often used in conjunction with topical drugs. The top layer of the dressing serves to protect the wound from infection, while the middle layer is usually made of absorbent material to maintain a moist environment and promote autolytic debridement, such as graphene-based materials. The bottom layer is composed of non-adherent material to prevent the dressing from sticking to newly granulating tissues. However, composite dressings can be more expensive and less flexible than other types of dressings [43].

### 2.6 Medicated Wound Dressings

Medicated wound dressings play an important role in the healing process by incorporating drug molecules into the dressing itself. After removing necrotic tissue, the use of antimicrobial chemi-cals can help to kill bacteria and promote tissue regeneration. One example of an antimicrobial dressing product is CutisorbTM. There are several types of silver-impregnated dressings available, including silicone gels, fibrous hydrocolloids, and polyurethane foam film. Additionally, the commonly used antiseptic iodine dressing works by interfering with bacterial protein activity and causing oxidative destruction of cells. Antimicrobial dressings are particularly useful in preventing and treating infections, es-pecially in the case of diabetic foot ulcers [44].

In conclusion, modern wound dressings not only provide protection for wounds, but can also be used for diagnosis and monitoring of the healing process. Recent efforts have focused on developing transparent wound dressings which offer several benefits, such as better visualization of the wound bed, assessment of healing progress, bacterial barrier, water-resistance, a moist environment, and promotion of selective debridement via autoly-sis [7,45]. The combination of transparency and antibacterial prop-erties make these dressings ideal for maintaining wound mois-ture and reducing infection rates. In contrast, traditional non-transparent dressings do not allow for monitoring of the heal-ing process and often require frequent changes, leading to an in-creased risk of infection. However, ongoing research aims to de-velop strong, transparent, long-lasting, and antimicrobial wound dressings [45].

## 3 Polymers for Fabrication of Transparent Wound Dressing

Modern wound dressings are based on synthetic, natural, and porous crystalline materials. [46–48].

### 3.1 Synthetic Polymers

The synthetic polymers in base dressings include polyurethane (PU) [49], poly(ethylene glycol) (PEG) [50,51], polyvinyl alcohol (PVA) [52], poly(lactic acid) (PLA) [53] and many others. Certainly, many synthetic polymers own some specific characteristics, such as mechanical integrity, good biodegradability profile, ease of surface modifications, thermal stability, biocompatibility, antimicro-bial features, capability to recreate a suitable environment for tis-sue regeneration, and their bioactivity that make them a suitable base material for the development of medical devices or dressings for clinical purposes [54–56]. Furthermore, some productive techniques allow fine-tuning of these properties. **Table 1** presents an overview of synthetic polymer-based wound dressings, with specific attention to their prominent characteristics.

Polyurethane (PU) has been widely used as potential semi-permeable wound dressing material because of its ability to provide a good barrier and permeability to oxygen, good biocompatibility, high mechanical strength, low cytotoxicity, and appropriate flexibility. It also evenly adheres to the wound and accelerates the epithelialization process [66–68]. PU nanofibrous membranes are considered skin substitutes; in addition, these membranes are impermeable to bacteria but are permeable to moisture vapor and control water vapor transmission [3].

Gholami et al. developed a novel polyurethane antibacte-rial dressing using carbonated soybean oil as an environmentally friendly, renewable resource-based raw material. Results showed good cytocompatibility, antimicrobial activity against various microbial strains, good tensile strength, preserve the moist environment, and a water vapor transmission rate of 390 g m$^2$day$^1$, In vivo assay on a rat [69]. In another work, Jatoi et al. synthe-sized polyurethane nanofibers-based composite containing silver nanoparticles/zinc oxide nanoparticles to improve the antibacte-rial characteristics of the wound dressing. The results confirmed that the PU-based composite nanofibers have exceptional bacte-ricidal and bacterial growth inhibition features that make them suitable for practical purposes in the healthcare system [70].

Gao et al. fabricated novel antibiotic delivery systems as efficient antibacterial healing material by using AgNPs-loaded N-[(2-hydroxy-3-trimethyl ammonium) propyl] chitosan/hyaluronic acid (HA) porous microspheres. The fabricated material effectively prevent the bacteria from invading and infecting the wound, accelerating the healing process of tissue regeneration at wound sites [71]. In another work, Luo et al. developed a vascular endothelial growth factor and Eumenitin co-loaded multi-functional methacrylated-k-Carrageenan using microfluidic elec-trospray for the treatment of chronic wound healing. Thus, ethacrylated-k-Carrageenan with sustained drug release proper-ties are able to serve as an effective platform to protect and de-liver bioactive proteins and small molecules to the wound surface for promoted regeneration [72].

Another synthetic polymer commonly used for the develop-ment of hydrogel-based dressings is poly (vinyl alcohol) (PVA) because of its exceptional biocompatibility, biodegradability, and showing good mechanical properties which makes it the right choice for various biomedical applications particularly wound dressing. Rathod et al. showed that the PVA hydrogel sheet incor-porated with calendula officinalis flower is a convenient and ef-fective wound dressing due to its anti-inflammatory, antioxidant, and anti-odemataous activity of the extract [63]. Similarly, Gao et al. introduced a green approach for the fabrication of electro-spun poly (vinyl alcohol) nanofibers incorporated with epidermal growth factors for wound healing applications. It was observed In vitro study that the fabricated dressings effectively promote cell proliferation [73].

### 3.2 Natural Polymers

In the past, a various natural material was utilized for the heal-ing process and wound closures due to their antioxidant and an-timicrobial properties, including propolis and honey [41]. Propolis is extracted from hives of honeybees and it was a proven fact that propolis is efficient for second-degree burn wounds. It has a better ability to achieve wound closure than silver sulfadiazine. Similarly, honey has an extended history in the healing process, treatment of burns, and ulcers since ancient times [74].

Natural materials have already attracted much attention in the biomedical field specifically in the field of wound dressings due to their many advantages, such as biocompatibility, biodegrad-ability, ability to recreate ECM architecture, triggers the healing process, and availability in abundance. Among natural bioma-terials, polysaccharides stand out in the field of wound healing due to their unique features such as active healing characteris-tics, accelerate regeneration and restoration of the impaired tis-sue, and providing a moist environment such as collagen [75], fib-rinogen [76], silk fibroin [77], cellulose [78], hyaluronic acid [79,80], algi-nate [81,82], chitosan [65,83,84], and others **Table 2**. In addition, nat-ural polysaccharides do not activate an adversative immunogenic reaction, but they interact with immune system components to accelerate the initiation of macrophages, which are key perform-ers in the wound care process [26,52].

Collagen is a natural protein biomaterial that is the most abun-dant protein in ECM proteins and the human body with excellent advantages such as bioactivity, biocompatibility, superb hemosta-sis, and favorable cell adhesion and proliferation [90]. Zhu et al. prepared adhesive collagen-based hydrogels with antibacterial ability using arginine and dopamine as modifiers. It exhibited that hydrogels' significant enhancement of antimicrobial activity, biocompatibility, and the promising interactions between func-tional groups of collagen molecules and arginine contributes to the enhanced gel characteristics [91]. In another study, Deng et al. designed a chitosan-based hydrogel that significantly accel-erates wound healing without adding therapeutic drugs. The re-sultant hydrogels demonstrated excellent self-healing properties, low swelling rate, good biocompatibility, accelerate cell prolifer-ation, hemostatic effect, and exceptional antibacterial activity [86]. Similarly, Wei et al. showed that the photo-induced adhesive car-boxymethyl chitosan-based hydrogels exhibit a good BSA adsorp-tion capacity, cytocompatibility, and hemostatic properties. The functional groups of additives endowed the hydrogel with inher-ent excellent antioxidant and antibacterial properties that enable hydrogels to promote wound healing efficacy [87].

### 3.3 Surface-Engineered Porous Crystalline Materials

Recently, a class of newly emerging porous crystalline polymers has gathered considerable attention for efficient wound dress-ing applications, in particular, nanoscale covalent organic frame-works (COFs) [93–95] and metal-organic frameworks (MOFs) [96–98]. These porous materials have highly ordered structures composed of organic ligands and inorganic metal ions with many appealing features, including high specific surface areas, high porosity, ex-cellent thermal stability, facile functionalization, good biocompat-ibility, and promising biodegradation ability [99,100]. This emerging class of porous multifunctional materials has potentialls in the biomedical field, such as drug delivery, high loading capacity, and loading of different small molecular drugs [101–106]. Li et al. de-veloped COFs-based wound dressings that inhibit bacterial infec-tions and accelerate wound healing. The resultant electrospun nanofiber-based wound dressing showed perfect physicochemi-cal properties including hydrophobic behavior high water uptake capacity, good biocompatibility, admirable antibacterial activity, and sustained release profile, which offer a safe environment for wound healing [107]. MOFs are porous crystalline structures com-prising metal ions (or metal clusters) connected through organic linkers. Their unique configurations made them special in the medical field because several biological practices are related to sorption. MOFs are widely used for wound healing owing to their high surface area (available for loading of cargos), easy surface modification (beneficial to biomedical applications) and tunable pore sizes (easy to wrap a variety of therapeutic drugs), and an-tibacterial properties as compared to other nanomaterials [108].

Chen et al. used MOFs alone to develop wound dressing with enhanced antibacterial activity for chronic wounds infected by multidrug-resistant bacteria. The resultant nanofibers-based wound dressing exhibits high biocompatibility and minimal cyto-toxicity [109]. Karakeçili et al. prepared a biocompatible scaffold a potential bone replacement and drug delivery system for the treatment of serious bone infections. These scaffolds were pre-pared by using a wet spinning technique and characterized to determine the morphology, swelling behavior, and antibacterial properties. In addition, a drug antibiotic vancomycin was loaded into ZIF8 and it was observed that the drug was released in a pH-controlled manner from the chitosan scaffolds. The final experi-mental results showed that not only antibacterial properties and bactericidal effects improved but ZIF8 incorporated scaffolds in-fluence positively the proliferation phase [110]. Another advantage of these porous materials (i.e., MOFs and COFs) are that they have been demonstrating exceptional potential as drug carrier in the biomedicine application.

## 4 Transparent Electrospun Nanofiber Dressings

Electrospun nanofiber-based dreesing materials with inherent an-timicrobial properties have drawn a lot of attention since they can be utilized to speed up wound healing by avoiding infection and dehydration. **Table 3**. shows the diverse antimicrobial electro-spun polymeric nanofibers as a wound dressing application. Cur-rently, accessible antimicrobial dressings made monitoring wound sites challenging. The principles of these techniques as well as ad-vances to develop transparent wound dressings are discussed in the following sections.

Electrospinning is a versatile manufacturing technique for the production of continuous ultrafine fibers with diameters ranging from tens of nanometres to several micrometers **Fig.**1(A). It is a simple and flexible technique for generating nanofibers structures, which are controlled by various parameters that are categorized into three main groups, (1) environmental parameters, (2) solution parameters, and (3) processing parameters.

Accumulated charge at the spinneret orifice exposes the discharged polymer solution to an electric field, causing a conically formed geometry like Taylor cone. As the electric field intensity grows, so does the accumulated charge on the surface of the bud-ding polymer droplet, resulting in repulsive electric forces that overcome the surface tension of the polymer solution. When the drawn polymer thread is directed to the grounded collector, the remaining solvent evaporates, keeping only the charged polymer fibre on the collector. The collector might be drum collector, rotating mandrel, tube collector, mandrel collector, frozen man-drel, thin disk collector, double-ring collector, and frame collector **Fig.**1(B) [130].

In a conventional configuration, only one needle with one thread of nanofiber is drawn from of the electrospinning setup **Fig.**1(C), Parallelizing the electrospinning of nanofibers from the spinneret is easy way to scale up production of nanofibers. Since each needle produces nanofibers through electrospinning, the production rate of nanofibers rises linearly with the number of needles. Other techniques have been developed to manufacture conical tips in the spinneret that are comparable to those produced by needles without the need of a mul-tiple of needles **Fig.**1(D). Using magnetic fluids as a template is one simplest way to produce sharp conical ends in the polymer solution reservoir. **Fig.**1(E) A coating of polymer solution on top of the magnetic field can mimic the morphology of the sharp con-ical tip and spinneret is form when a high voltage is put between conical tip and the collector. Another popular method for increas-ing the production of nanofibers is to use a rotating cylinder with a rough surface to dynamically generate sharp conical points, as shown in **Fig.**1(F).

These nanofibers exhibit tuneable pore size and morphology, high porosity, and excellent mechanical properties that can make them favorable or waterproof breathable wound dressings and other biomedical applications such as tissue engineering, antibacterial dressings, and drug delivery systems [131–139]. **Table 4** presented experimental parameters for the fabrication of transparent nanofibers by using electrospinning techniques.

Despite all the advantages of nanofiber-based dressings, accu-rate wound assessment, and effective wound management re-quire an understanding of the physiology of wound healing. The electrospun nanofiber mats are inherently non-transparent but various techniques have been employed to develop trans-parent nanofibers mats for different applications such as air fil-ters, masks, and other biomedical applications specifically wound dressings [145].

An advanced biomedical application of transparent nanofibers is to replicate the eye cornea. Generally, the electrospin-ning method is to fabricate two-dimensional (2D) structured fi-brous. Kim et al. advised a novel electrospinning method with some modifications to develop a 3D hemispherical transparent nanofibers scaffold that mimics the real properties of the cornea. Results suggested that this modified electrospinning technique not only produces hemispherical scaffolds such as contact lenses, but it can design curved tissues such as those of the eye, elbow, or 3D wound area [140]. Zhang et al. fabricated the most effi-cient, transparent, rubbery, and multifunctional 2D electrospun air filters from poly (mphenylene isophthalamide)/polyurethane (PMIA/PU). These 2D electrospun filters have a superlight base weight (0.12 gm$^{-2}$), low thickness (350 nm), self-sustaining abil-ity, good optical transparency (~ 95.0 %), and high bioprotec-tive properties to filtrate particulate matter 2.5 with > 99.9. effi-ciency [146].

Wang et. al designed high light-transmitting fibrous by optimizing a self-designed topological frame structure with high light flux [147]. The procedure for creating transparent light-transmitting fibrous membranes (HLTMs) is shown in **Fig.**2(A-B). Electrospun fibers were directly deposited on a mesh substrate, such as flat aluminum foil (S1), rectangular mesh (S2), hexagonal mesh (S3), and outer hexagonal frame with a middle herringbone frame (S4). The membranes fabricated on substrates S1, S2, S3, and S4 exhibited light transmittances of 50 %, 63 %, 61 %, and 80 %, respectively. It was discovered that the fibrous structures fabricated with S1 offered randomly oriented and disorganized 3D networks with very less light transmittance **Fig.**2(C), In the case of S2 and S3, the fibers were mainly piled and tightly packed in the middle of the mesh, these packed fibers deflect the light rays resulting in low light transmittance **Fig.**2(D-E). However, in the fibrous mem-brane with S4, the majority of fibers were deposited mostly along the hexagonal and herringbone frames. In the middle regions be-tween the hexagonal and herringbone frames, there were a small number of fibers deposited. This specific fiber distribution per-mits more light to pass through the membrane from the center of three areas with minimum light loss **Fig.**2(F). As shown in the FESEM image, three different regions in S4 fibrous membrane, the first part was the outer hexagonal frame, the second part was the herringbone frame in the middle, and the third part was the three regions between the hexagonal and herringbone frame **Fig.**2(G) [147]. The thickness and porosity of fibrous membranes are two key parameters that influence transparency. The 90 %, 84 %, 78 %, and 62 % transparency of the S4 membrane were predominantly concentrated with the pore sizes 2.6-2.8µm, 2.3-2.5µm, 1.5-1.8µm, and 1.0-1.3µm, respectively **Fig.**2 (G). These HLTMs were used as face masks. But the fabrication strategy may be also useful for developing transparent wound dressing.

Ma et al. developed a biocompatible electrospun PCL/shellac/PCL sandwich-structured nanofiber mat loaded with salicylic acid as a guest drug [141]. A graphic illustration of a PCL/shellac/PCL sandwich structured membrane as shown in **Fig.**3 (A). During the ethanol treatment process, shellac nanofibers melted and filled the pores of the PCL nanofibers until a uniform coating formed on the membrane surface. After applying the ethanol vapor treatment, the membranes show excellent transparency with good smoother properties. The surface morphologies of the PCL/shellac/PCL nanofiber membranes before and after treatment are presented in Fig **Fig.**3 (B-C) respectively. Results showed that ethanol vapor-treated membranes retained good mechanical properties, 8 h-drug release profiles, and transparency. **Fig.**3. These characteristics satisfy the demands for use of the membranes in skin care applications on the face. (D) shows the surface morphologies of the PCL/shellac/PCL nanofiber membranes both before and after treatment [141]. Similarly, Wang et al. fabricated ultrathin transparent nanofibers from silkworm silk through electrospin-ning and carbonization techniques, then integrated with planar polydimethylsiloxane films to construct highly sensitive skin-like pressure sensors. These artificial pressure sensors have prime importance in different applications such as health monitoring, disease diagnostics, and smart robots. **Fig.**3 (D) demonstrates the preparation process of Carbonized silk nanofiber membranes (CSilkNM) pressure sensors. The CSilkNM-30 pressure sensors have good optical transparency and flexibility, **Fig.**3 (E-G) shows the interconnected structure of nanofibers and indicates that the fibrous morphology sustained the high-temperature pyrolysis treatment respectively. Results exhibit the successfully fabricated of the flexible, transparent, and ultrathin CSilkNM on large scale from silkworm silk through a simple electrospinning and heat treatment process for electronic skin application [142].

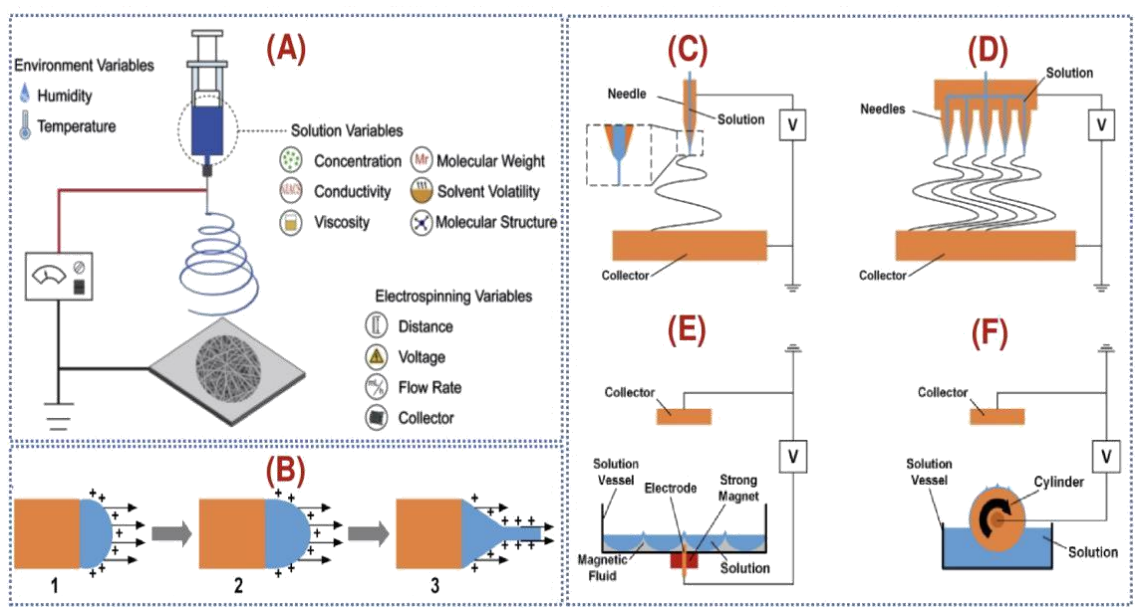

Fig. 1 Schematic diagram of lab-scale electrospinning setup [129]. (Adapted with permission from ref [129], Copyright 2017, Wiley-VCH Verlag GmbH & Co. KGaA, Weinheim, Germany.) (B) Formation of the Taylor cone under electric field. When the electric force that pulls the polymer solution downstream is strong enough to overcome the surface tension that holds the polymer solution to the tip, a tiny jet of the polymer solution is drawn from the tip of the Taylor cone and its solvent evaporates quickly in air, forming solidified polymer fibers on the collector, Scale-up of electrospun nanofibers: (C) A typical electrospinning system with a single needle in the spinneret, (D) Parallel scale-up of nanofibers drawn from an array of needles in the spinneret, (E) Instead of using needles, numerous static sharp conical tips are built up in the polymer solution reservoir by using a magnetic fluid as template, (F) Dynamic generation of multiple sharp conical tips through a rough rotating cylinder [130]. (Adapted with permission from ref [130], Copyright 2018, De Gruyter Open Access, Berlin, Germany.)

Recently, Our latest research paper has been published on the fabrication of transparent composite nanofibrous membrane. We used polyurethane (PU) to prepare the nanofibers mat, followed by reinforcing fibers in a polymer solution such as polymethyl methacrylate (PMMA), and polydimethylsiloxane (PDMS) to investigated the effect of the concentration and the immersion time on the transparency of the membrane **Fig.**3 (H). Both polymeric solutions improve the transparency of polyurethane nanofibrous membrane (PU/PMMA= 80%T; PU/PDMS= 5%T) while maintaining their mechanical strength with controlled solvent evaporation after impregnation. However, this work provides opportunities to develop transparent, breathable, and high-performance nanofiber-based membranes for targeted applications. [49].

A hydrophobic monohydrochloride monohydrate-free ciprofloxacin antibiotic was incorporated into a PVP bio-polymer and fully transparent antimicrobial films were developed by casting from aqueous solutions. The SEM morphology image of **Fig.**4 (A) shows the microscopic image of the developed mats. Fibers were generated quite consistently, as exhibited, with no beads and shows excellent antibacterial activities against Escherichia coli and Bacillus subtilis bacteria. The fabricated Films and nanofiber mats had promising wound resorption characteristics confirmed by in vivo full-thickness excisional skin wound healing mice model. Nanofiber mats were resorbed much faster than transparent films. Wound exudate absorption in the films and the resorption rate of the nanofiber mats were dependent on aqueous acetic acid concentrations. As such, these PVP/Cipro solutions in aqueous acetic acid can be used either to produce transparent soft films or nanofiber mats [143]. Xia et al. showed transparent antibacterial chitosan-coated cellulose membrane (CM-CS) for visualized wound healing dressing **Fig.**4 **(B)**. Both the key molecules were mainly combined by the intermolecular hydrogen bonding network and showed excellent tensile strength with a higher swelling ratio and porosity. The antibacterial test revealed that CM-CS had excellent antibacterial activity against S. aureus and E. coli. More importantly, in vivo animal experiment and histological examination revealed that CM-CS could be used as potential wound dressings to promote wound healing process **Fig.**4 (C) [144].

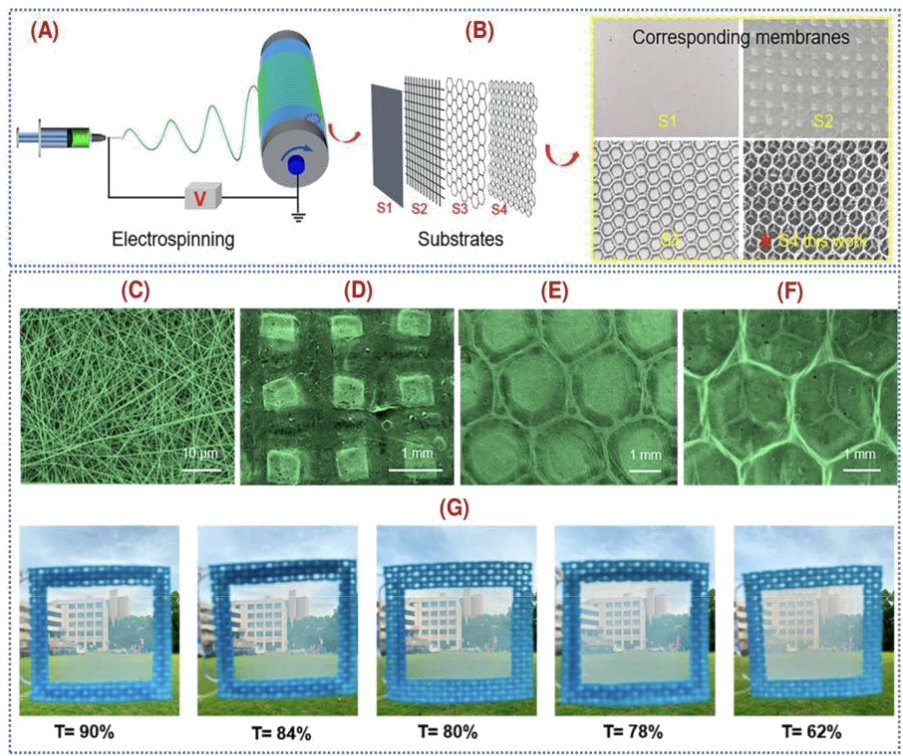

Fig. 2 (A) Schematic showing the fabrication of HLTFMs (Substrates of S1 to S4 are aluminum foil, rectangular mesh, hexagonal mesh, and self-designed mesh), (B) Optical images of fibrous membranes corresponding with S1 to S4, (C-F) FE-SEM images of fibrous membranes exhibited with S1, S2, S3, and S4 respectively, (G) Digital images of HLTFMs with different transmittance 90 %, 84 %, 80 %, 78 %, 62 % from left to right, respectively [147]. (Adapted with permission from ref [147], Copyright 2022, ACS, New York, United States.)

## 5 Transparent Composite Films/Membranes

Another effective way of preparing wound dressing is making a composite films/membrane. Generally, a single dressing do not possess all of ideal properties, but blending with additives or natural polymers can be improve the dressing material proper-ties. Extensive research has been conducted to fabricate compos-ite film/membranes as efficient wound dressings [85,148–156]. Patil et al. examined a biocompatible and biodegradable novel dress-ing by blending zinc oxide nanoparticles embedded in a silk fi-broin–polyvinyl alcohol (SF–PVA/ZnO) composite film. The com-posite film showed excellent mechanical and antibacterial prop-erties due to ZnO NPs [156].

Aritra Das et al. fabricated transparent Polyvinyl alcohol (PVA)/starch(St)/citric acid/glycerol (Gl) based composite films for wound dressing applications. The hydrogen network between -C=O and -OH played a major role in the formation of the trans-parent film. A PVA film was created with 2 g and 3 g CA con-centrations against E. coli, which enable the formation of a cir-cular clear zone of inhibition against L. monocytogenes **Fig.**5 (A-B). The PVA composite film was prepared with E. coli at 50 °C crosslinking temperature and 2 g CA concentration showed good results in **Fig.**5 (C) [157]. Recently, Delavari et al. devel-oped a PVA/Starch-based composite film with a special feature of biodegradability and transparency. Other than transparency these dressing presents improved water vapor transmission rate and an-tibacterial activity against the bacterial flora (various bacteria ex-istent in the air) [164]. By using an enzyme-linked immunosorbent assay to measure the cytokine concentrations in mouse serum, the effects of paramylon were also investigated by Yasuda et al. In this work, mice were treated with a paramylon film dressing and observed wound contraction. The wound area was collected on days 0 and 5 and observed that animals treated with paramylon film had more wound contraction than mice treated with ordinary cellulose after 5 days [165].

Wang et al. designed a highly transparent BC membrane and polyhexamethylene biguanidine bacterial cellulose (PHMB-BC) based composite membrane for wound healing applications. As shown in **Fig.**5(D), PHMB-BC membranes are significantly more transparent than BC membranes, so would allow for real-time vi-sualization of wound healing progress. **Fig.**5 (E-G) displays the macroscopic tear strength of various materials. Each material de-veloped a 5 mm crack. The BC membrane fully tore up due to the weight's gravity (550 g). The size of the cracks remained un-changed, and both materials had successfully withstood the force of the tear [158].

In another work, Huang et al. prepared a waterproof and breathable bandage with plasma-treated PE films as the hy-drophobic layer and a blend of polyvinyl alcohol–alginic acid by electronic spinning technique as a hydrophilic contact layer. **Fig.**6 **(A)** illustrated the fabrication method and how it works efficiently. These fabricated dressings have shown exceptional cooling ability, improved waterproof and breathable properties, and improved wound healing [159]. Zhang et al. developed cellulose/maghemite transparent nanocomposite membranes with excellent porosity with good adsorption capacity, excellent biocompatibility, and controlled release of doxorubicin as a potential wound dressing materials **Fig.**6 (B-C) [160].

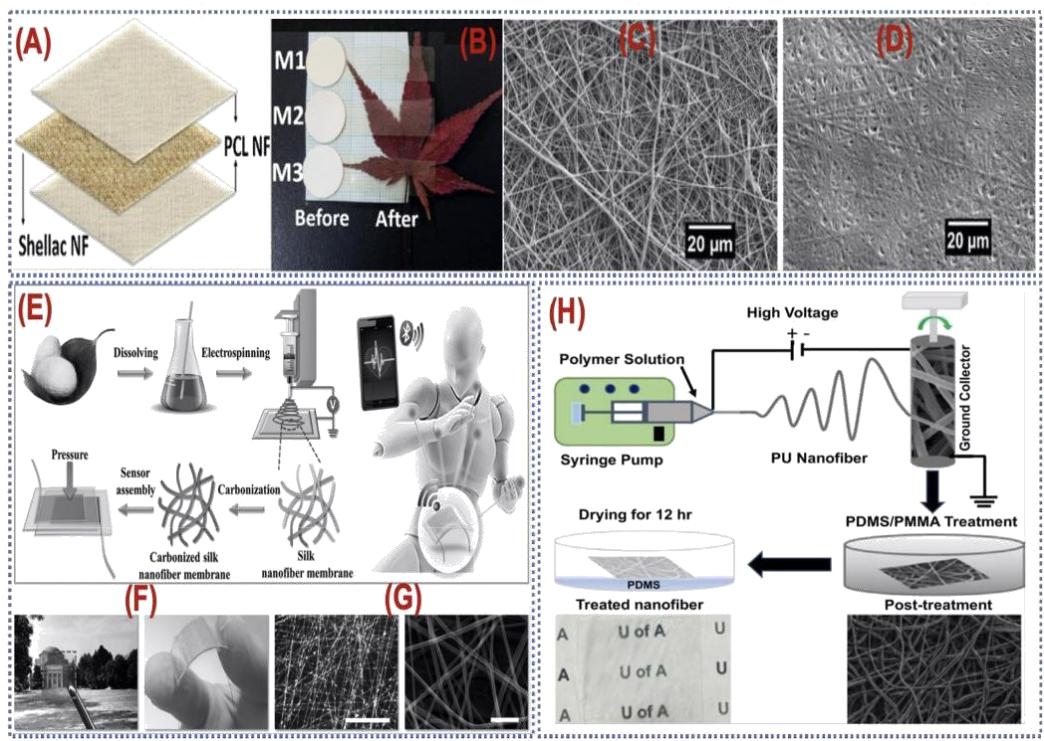

Fig. 3 (A) Illustrated schematic diagram of the sandwich structured membrane, (B) photographs of the prepared membranes before and after Ethanol vapor treatment (M1 & M3: top and down a layer of PCL membrane, M2: shellac membrane between M1 & M3), SEM images of PCL/shellac/PCL membranes (C) before (D) after treatment [141]. (Adapted with permission from ref [141], Copyright 2018, Springer, New York, United States.) Fabrication process and structure of the CSilkNM pressure sensor (E) Schematic illustration showing the fabrication process of CSilkNM pressure sensors, Photographs showing the (F) transparency and flexibility of the obtained sensor, (G) Optical and SEM image of a silk nanofiber membrane [142]. (Adapted with permission from ref [142], Copyright 2017, Wiley-VCH Verlag GmbH & Co. KGaA, Weinheim, Germany.) (H) Schematic diagram of electrospinning and post-treatment process [49]. (Adapted with permission from ref [49], Copyright 2023, Elsevier, Amsterdam, Netherlands.)

Robust, ultra-smooth, and flexible composite films of pure silk fibroin and silk blended with Polyvinyl Alcohol (Silk/PVA) have shown high optical transparency above 85 % as presented in **Fig.**6 (D). Moreover, these regenerated silk fibroin films offer magnificent mechanical, optical, and electrical properties useful for the development of bio-integrated electronic devices [162]. Kim et al. developed a tympanic membrane perforation using a nanofibril-lar bacterial cellulose (BC) patch. The BC nano fibrillar patch was transparent and it might be possible to observe the tympanic membrane as it regenerates. Images taken with a scanning electron microscope revealed that the BC nano fibrillar patch included nanoscale filamentous networks with an extracellular matrix-like structure **Fig.**6 (E-F). They conducted an In vivo animal study to investigate the BC nano fibrillar wound-healing application with tympanic membrane regeneration. Another advantage of this method was the amount of tympanic membrane regeneration is visible when the BC nano fibrillar patch was placed on the perforated tympanic membrane. It is challenging to compare the paper patch's ability to promote tympanic membrane healing to that of bacterial cellulose [161].

Similarly, Samadi et al. developed a transparent hydrogel film-based antibacterial wound dressing incorporated with silver nanoparticles [166]. These dressings have some favorable features including, moisture retention properties, absorb wound exudate, and prevent wound infection, due to the antibacterial properties of AgNPs. In addition, with good biocompatibility, and tissue regeneration, these dressings are promising for wound healing [166]. In another work, Ghiasi et al. fabricated a new generation of bio-functional Persian gum-based (PG) films via the co-delivery of crocin and cinnamaldehyde [163]. The optical, mechanical, and microstructural attributes of the films were more affected by the incorporation of the free form of bioactives. The UV–Vis light barrier property and hydrophobicity of the film were improved after incorporating single (SE) and double emulsions (DE) **Fig.**6 (G-J). Film-forming solutions of PG-SE and PG-DE originally pre-sented lower color intensities as the bioactives were encapsulated in emulsion structures before being added into the film formula-tions. However, the greater stability of bioactives within double emulsion against different pH (2 and 7) and temperatures (25 and 75 °C) produced more superior antibacterial and antioxidant activities **Fig.**6 (J). Furthermore, the homogenous distribution of double emulsion in the film matrix led to more resistant and ex-tensible films [163].

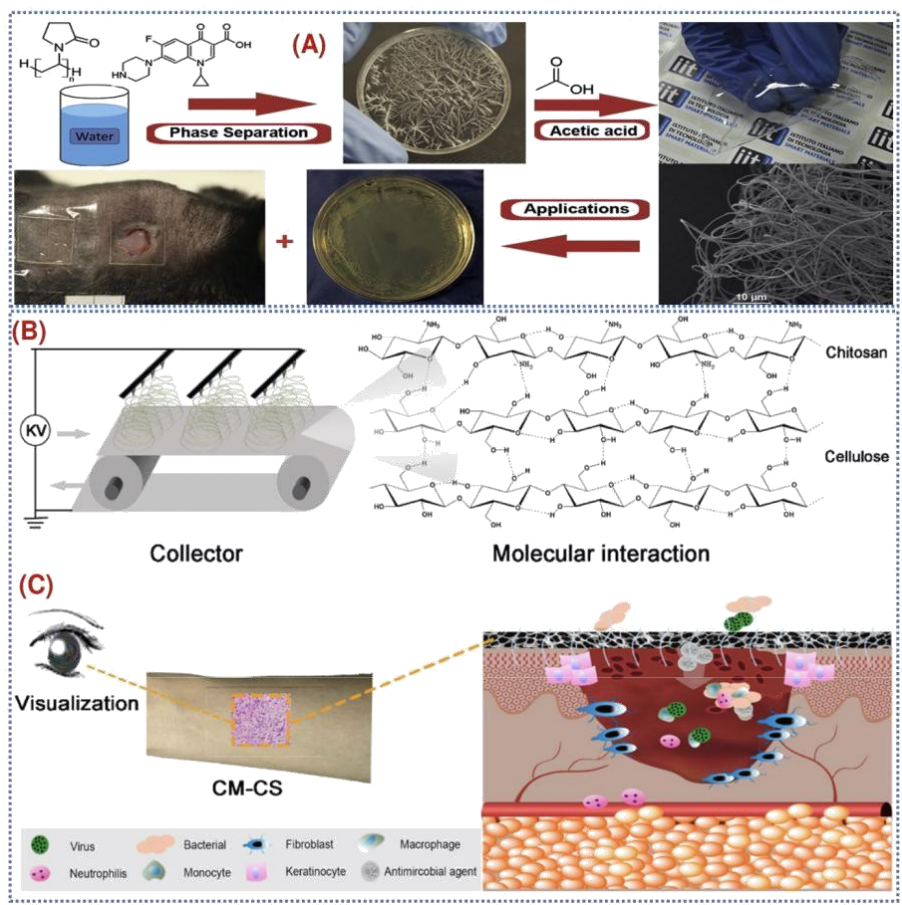

Fig. 4 (A) Photograph of nanofiber mate and morphology of the fibers [143]. (Adapted with permission from ref [143], Copyright 2017, Elsevier, Amsterdam, Netherlands.) (B-C) Schematic Illustration of the preparation of CM-CS and CM-CS Covering the Skin Wound [144]. (Adapted with permission from ref [144], Copyright 2020, ACS, New York, United States.)

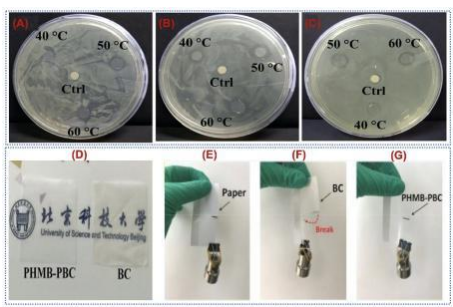

Fig. 5 (A-B) Antibacterial activity against L. monocytogenes with 2gm and 3gm CA, and (C) E. coli of PVA/St films prepared with 2 gm CA con-centration [157]. (Adapted with permission from ref [157], Copyright 2019, Elsevier, Amsterdam, Netherlands.) (D) Photograph of nanofiber mate and morphology of the fibers, (E-G) The macroscopic tear strength of paper, BC and PHMB-PBC [158]. (Adapted with permission from ref [158], Copyright 2019, Elsevier, Amsterdam, Netherlands.)

# 6 Transparent Crosslinked Hydrogel Wound Dress-ings

Wound dressings have been developed by using hydrogel [167–172], hydrocolloids, foams, and scaffolds. Among all, hydrogels represent a class of materials that are most commonly utilized for the management of acute-to-chronic wounds and tissue engineer-ing applications. Hydrogels are three-dimensional (3D) networks composed of hydrophilic polymers that are physically or chem-ically crosslinked [173–178] The cross-linked structure of polymer chains exhibits the potential to absorb a large amount of wound exudates and hold it while separating the bacteria, and odorous molecules from the exudate. In addition, hydrogel as a wound dressing and its high aqueous content permit oxygen into the wound to excite the wound healing process, providing a moist environment, a comforting effect to the wound, and a controlled drug release profile [179,180]

Li et al. proposed a strategy to develop novel biomechani-cally active and biochemically functional hydrogel dressings to help wound closure and promote wound healing. These dress-ings are biomechanically active with excellent features including thermo-responsive self-contraction. At the same time, these dress-ings demonstrate self-healing features, temperature-dependent drug release profile, anti-infection, antioxidation, and conductiv-ity. **Fig.7** (A-B) exhibits various dressings and their corresponding multifunctional properties that significantly accelerate the wound healing process. The results showed hydrogel dressing presents good swelling ratios, similar conductivity to the human dermis, good biocompatibility, antibacterial ability, antioxidant capacity, and helps closure of wounds [181].

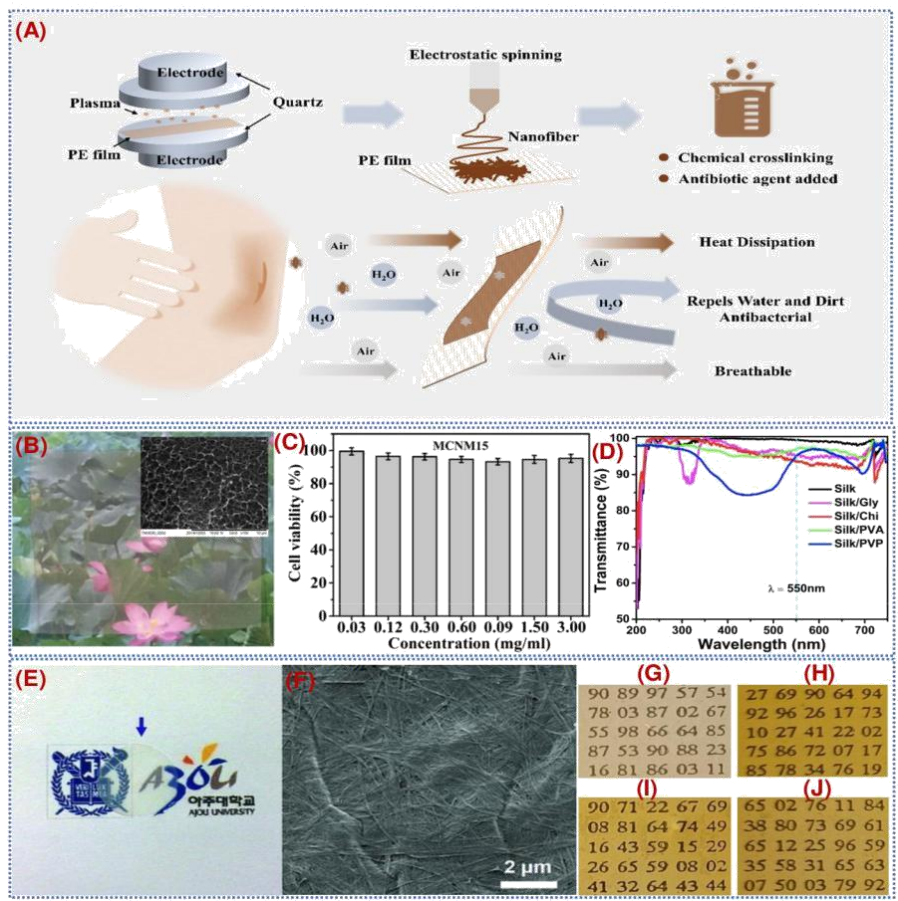

Fig. 6 (A) Illustration of the fabrication process of composite dressings and their characteristics [159].(Adapted with permission from ref [159], Copyright 2022, Springer, New York, United States.) (B-C) Photographs and SEM images of the surface of the maghemite/cellulose nanocomposite membranes [160]. (Adapted with permission from ref [160], Copyright 2017, Elsevier, Amsterdam, Netherlands.) (D) Transparency of prepared films measured via UV–Visible spectroscopy [161]. (Adapted with permission from ref [161], Copyright 2013, Wiley-VCH Verlag GmbH & Co. KGaA, Weinheim, Germany.) Representative (E) digital camera image, (F) SEM image of the BC nano fibrillar patch [162]. (Adapted with permission from ref [162], Copyright 2021, Elsevier, Amsterdam, Netherlands.) Visual appearance of (G) control Persian gum (PG)-based edible films and those incorporated with crocin and cinnamaldehyde in (H) free and encapsulated form within (I) single (SE), and (J) double (DE) emulsions [163]. (Adapted with permission from ref [163], Copyright 2023, Elsevier, Amsterdam, Netherlands.)

Despite the various advantages of photothermal hydrogel dressing for wound treatments, still, a major challenge that limits their clinical applications in the healthcare system is their poor transparency due to the high absorption of incident light [182,183]. Transparent hydrogels as a wound dressing is an encourag-ing technology for wound treatment applications and developed a composite photothermal hydrogel dressing with high trans-parency. Results exhibit that successfully a hydrogel wound dress-ing with high transparency and accelerated healing has been de-veloped and it shows excellent photothermal conversion ability and high transparency [182].

Chen et al. developed self-healable and injectable hydrogel from the multi-arm thiolated polyethylene glycol (SH-PEG) with silver nitrate (AgNO3) **Fig.**8(A-B). Due to the encapsulation of the desferrioxamine drug into the 3D hydrogel, they obtained a multi-functional hydrogel with antibacterial, and angiogenic properties, subsequently accelerating tissue healing in diabetic skin wound sites **Fig.**8 (C-D) [184]. With a focus on the effectiveness of wound healing in response to mechanical softness/stiffness, Huacheng et al. fabricated zwitterionic poly (sulfobetaine methacrylate) (PSBMA) transparent hydrogels. The elastic modulus ranges from a softness of 10 kPa to a hardness of 60 kPa and is used to full-thickness excisional acute wound regeneration in mice **Fig.**8 (E). Overall findings demonstrated that the softer PSBMA hydrogels may effectively accelerate wound healing through the intrinsic elastic impulse to enhance neovascularization [185].

Similarly, Zhang et al. fabricated a transparent and biocompati-ble hydrogel dressing with sericin and polyacrylamide. These pre-pared hydrogel dressings possess biodegradation ability, porous structures good swelling behaviors, and mechanical strength that can be controlled by regulating the content of sericin. More-over, the hydrogel system is well-suited to hosting cells due to its superb cell-adhesive ability, which efficiently accelerates cell attachment, proliferation, and long-term stability. These re-sults indicate that the sericin-polyacrylamide IPN hydrogel can be used as a visualized dressing material for real-time monitoring of wounds.[186]. In other studies, Thet et al. first time created a pro-totype "intelligent" wound dressing, that was triggered by contact with bacterial biofilms in several model settings **Fig.**8 (F). In the first experiment, produced hydrogel film was composed of vesi-cles and agarose. Further developments were achieved by pat-terning vesicles in a regular array of wells, which upon release offers a rapid dilution of fluorescence into the agarose **Fig.**8 (G). The dressing activating was confirmed by lysing vesicles with the addition of Triton detergent and showing the "activation" of fluo-rescence as well as the clear and uniform signal created over the whole dressing **Fig.**8(H) [187].

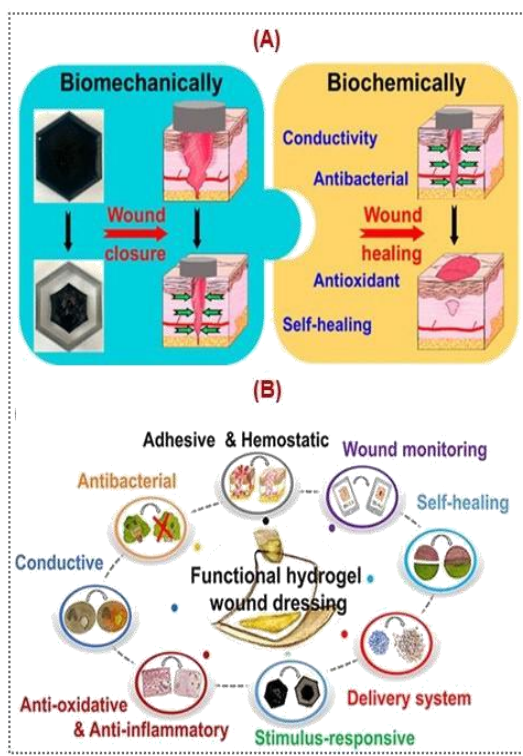

Fig. 7 (A) Presents how dressings help biomechanically and biochem-ically in healing [181], (Adapted with permission from ref [181], Copyright 2020, ACS, New York, United States.) (B) shows hydrogel wound dress-ings and their characteristics [167]. (Adapted with permission from ref liang2021functional [167], Copyright 2021, ACS, New York, United States.)

Jun Li et al. oxidized hydroxyethyl starch (O-HES) and mod-ified carboxymethyl chitosan (M-CMCS) to design the in-situ forming hydrogel with outstanding self-recoverable extensibility-compressibility, biocompatibility, biodegradability, and trans-parency for expediting wound healing **Fig.**8 (I). The degree of O-HES oxidation and M-CMCS amino modification was 74.0 % and 63.0 %. Sprague-Dawley rats with full-thickness skin defects were used in experimental research on the In vitro healing of skin wounds **Fig.**8 (J). Significant results were obtained such as in-creased granulation tissue production, accelerated epithelializa-tion, and reduced collagen deposition seen in the group treated with M-CMCS/O-HES hydrogel [188]. A polyurethane urea/poly(N-isopropylacrylamide) based thermo-sensitive and transparent bio-compatible hydrogel showing excellent wound healing with cell adhesion control properties. Even at very low input compositions of NiPAAm, 3D polymer networks developed by using a thermal free radical polymerization process **Fig.**9. These smart materials show potential applications for cell transplantation in the wound-healing process [189].

Tamayol et al. designed pH-responsive hydrogel for epidermal applications. The size of the manufactured pH-responsive hy-drogel fibers was controlled by utilizing a microfluidic spinning technique to create the alginate-based microfibers **Fig.**10 (A-B). The fabricated materials were applied on explanted pig skin, and agarose gel was sprayed with various pH solutions (pH = 6.2, 7.2, and 8.2) **Fig.**10 (C). After 30 minutes, when the fiber col-ors were stable, pictures were taken using a smartphone. The developed skin dressing can be used healthcare tool to track the progress of a wound's healing process [190]. Shao et al. synthe-sized tannic acid, thioctic acid, and phytic acid (TATAPA) based hydrogel via a bottom-up method. The hydrogel showed ad-hesive, non-toxic, transparent, hemostasis, and free from toxic crosslinkers, oxidants, or heavy metal ions. Therefore, the TAT-APA hydrogel can be employed as a first-aid bandage for superfi-cial lesions **Fig.**10 (D) [191]. A hydrazide-modified hyaluronic acid and benzaldehyde terminated F127 triblock copolymers to de-velop the novel a novel hydrogel (HA-az-F127 hydrogel) **Fig.**10 (E). Additionally, the hydrogel has unique properties, such as ad-justable mechanical strength, self-healing ability, liquid absorp-tion or drainage, tissue adhesion, and effectively promoting of burn wound repair [192].

Wei Yang et al. developed the cysteine-containing ultrashort peptides transparent hydrogel for wound healing applications **Fig.**10 **(F**-I). During the gelation, the disulfide bond accounted for the great shape fidelity of crosslinked gels even after extended du-rations of submersion in water and held the fibers together **Fig.**10 (I). The crosslinked gels provide an advantage over other non-crosslinked peptide hydrogels for burn wounds because they can be easily handled with forceps during surgical manipulation This peptide hydrogel formulation was safe for topical use and pro-moted a faster, better-quality wound healing process [193].

Tonsomboon et al. demonstrated transparent membranes from common and natural polymers: gelatin and alginate. As soon as the gelatin mats came into contact with the alginate solu-tion, they shrunk and transformed into transparent film, these films were crosslinked in calcium chloride ($CaCl_2$) solution to produce stiff hydrogels. Using either aligned or randomly ori-ented gelatin mats, the produced hydrogels were transparent. Depending on the thickness of the gelatin mats, the hydrogels' thicknesses ranged from 0.4 to 0.6 mm. According to the ratio of dry gelatin mat to hydrated composite (n = 10), the compos-ite hydrogels from randomly oriented fiber include approximately 16 wt % of gelatin fibers, whereas those made from aligned fibers have roughly 15 wt %. In comparison to a homogeneous algi-nate hydrogel (without gelatin fibers) and porcine cornea, the transparency of composite hydrogels comprising various types of randomly oriented gelatin fibers. The transparency of all hydrogels with the same thickness as the human cornea (0.5–0.6 mm) and porcine cornea, which is roughly twice as thick as the created hydrogels. The homogenous hydrogels are capable of transmit-ting the majority of visible light. Compared to the pig cornea, gelatin fibers can transmit a somewhat higher percentage of light whether they are water crosslinked or not. However, hardly any light was transmitted through the ethanol-crosslinked gelatin fiber-containing alginate hydrogels shown in **Fig.**10 (J). [194]. Fi-nally. These novel nanofibrous composites have great promise as scaffolds for corneal tissue engineering applications due in part to their higher optical transparency.

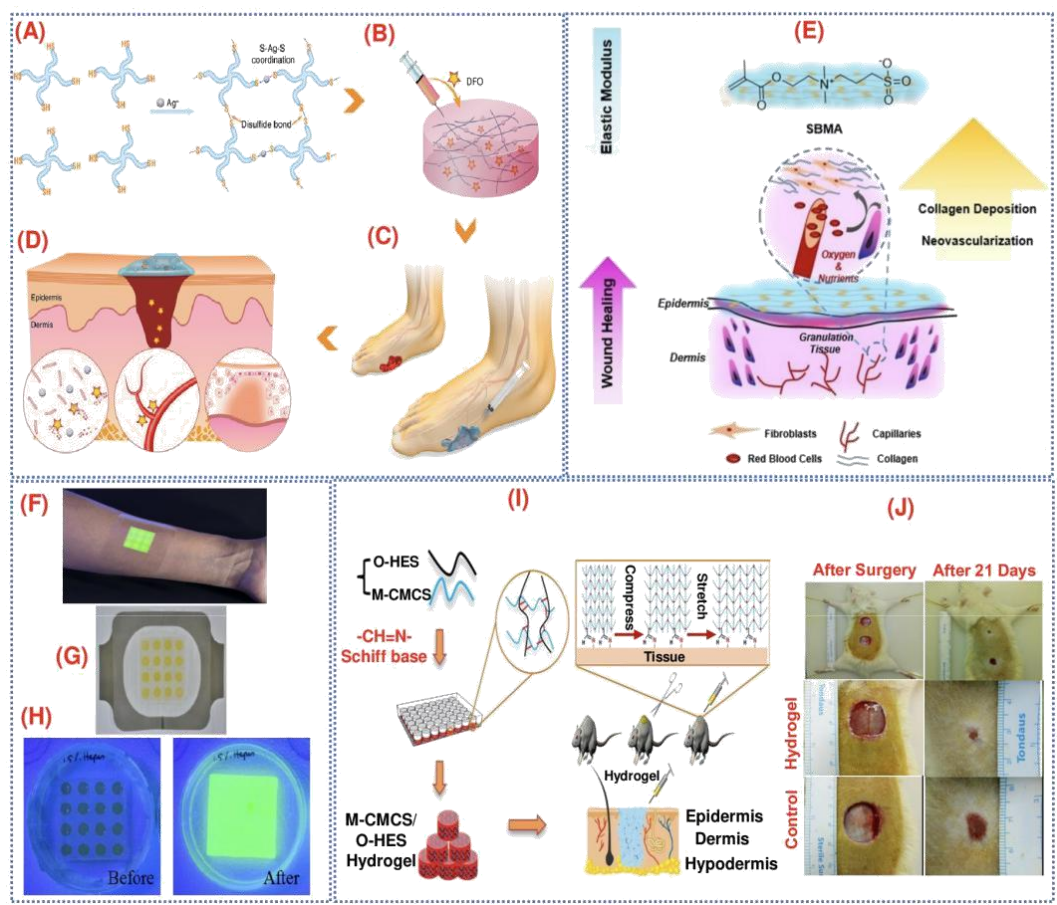

Fig. 8 (A) Schematic image of the self-healing Ag(I)-thiol (Au–S) coordinative hydrogel developed by mixing 4-arm-PEG-SH with AgNO3, (B) In situ encapsulation of drug (DFO) to obtain an injectable, self-healing, antibacterial, and angiogenic multifunctional hydrogel for diabetic skin wound repair, (C) Foot ulcers of type I diabetes (left) and therapeutic effect after hydrogel treatment (right), (D) Mechanism of the hydrogel in repairing skin defects through injection [184]. (Adapted with permission from ref [184], Copyright 2019, Springer Nature, Berlin, Germany.) (E) Zwitterionic poly(sulfobetaine methacrylate) hydrogels for wound healing [185]. (Adapted with permission from ref [185], Copyright 2019, Royal Society of Chemistry, London, United Kingdom.) (F) Design and development of the intelligent hydrogel wound dressing, (G) a finished prototype dressing on a transparent polypropylene film, (H) the same dressing before and after the activation using Triton to be seen under UV light [187]. (Adapted with permission from ref [187], Copyright 2016, ACS, New York, United States.) (I) Schematic image of in situ forming hydrogel formation from oxidized hydroxyethyl starch (O-HES) and modified carboxymethyl chitosan (M-CMCS) with outstanding self-recoverable extensibility-compressibility, biocompatibility, biodegradability, and transparency for expediting wound healing, (J) Representative images of full-thickness skin defects treated with M-CMCS/O-HES hydrogel or not on days 0, and 21 after surgery [188]. (Adapted with permission from ref [188], Copyright 2020, ACS, New York, United States.)

Recently, Seshadri et al. reported a novel and smart wound dressing using polyvinyl alcohol (PVA) as the base material with some excellent features including, absorbent, flexible, transpar-ent, and inexpensive moisture-management (AFTIDerm). Vari-ous concentration of GI was added to prepare PVA/GI dressings and GI 5 % was observed as an optimized concentration. Results showed that the resulting dressing exhibits good biocompatibil-ity, absorptive abilities, and scalability, and additionally PVA/GI substrate material expresses the potential for smart dressings. The wound dressing at optimized concentration shows relatively good stability in absorption compared to lower concentration over one week; moreover, this methodology to prepare AFTIDerm is amenable to addressing the diversity of clinical wounds [9]. Clin-ical treatment faces a challenging issue with treating and man-aging diabetic foot ulcers. A transparent monitoring system was used to treat and control diabetic foot ulcers with blood seeping and hard-healing, which presented challenges for the other non-transparent conductive patches. **Fig.**10 (K). The poly(tannic acid) doped polypyrrole nanofibrils in the poly(acrylamide-acrylate adenine) polymer networks based transparent hydrogel acceler-ates hemostasis, enhance cell-to-cell contact, prevent wound in-fection, facilitate collagen deposition, and stimulates angiogen-esis to visibly and efficiently aid the healing of diabetic foot ulcers [195].

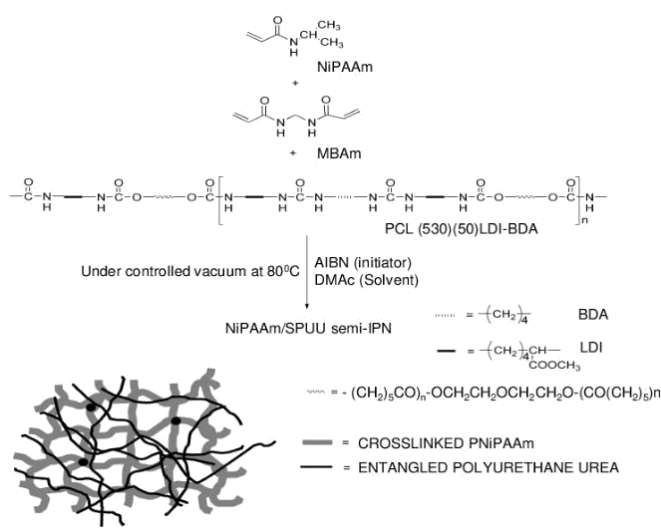

Fig. 9 Schematic diagram showing components used for the synthe-sis and the resulting structure of thermo-responsive transparent semi-interpenetration polymer networks [189]. (Adapted with permission from ref [189], Copyright 2008, ACS, New York, United States.)

## 7 Concluding Remarks: Current Challenges and Fu-ture Prospects

Wound healing is an intricate and dynamic process, requiring the right environment for each phase. As such, the selection of an appropriate dressing material for specific injuries must be considered when fabricating a dressing. A wide range of nat-ural and synthetic polymer dressings are available to meet the requirements of transparency, wound healing, antibacterial, and other properties of the dressing. This review paper provides an overview of the most recent studies on transparent wound dress-ings, with a particular focus on the electrospinning method for the preparation of various types of transparent wound dressings. Furthermore, new techniques for the fabrication of hydrogels have been highlighted as potential avenues for the development of transparent dressings. Further studies are needed to evaluate and develop innovative approaches to prepare the transparent film/nanofibers-based for wound dressing purposes. For exam-ple, preparing bi-layer transparent wound dressings with dynamic and intelligent drug carriers, constant drug release profiles, and multi-functionality dressing remains to be solved.

## List of abbreviations

COFs: Covalent organic frameworks
  ECM: Extracellular Matrix
  MOFs: Metal-organic frameworks
  CM-CS: Chitosan-coated cellulose membrane
  PCL: Polycaprolactone
  PU: Polyurethane
  PLA: Polylactic acid
  PVP: Polyvinylpyrrolidone
  PVA: Poly (vinyl alcohol)

PEO: Polyethylene oxide
PLGA: Poly (lactic-co-glycolic acid)
PEG: Polyethylene glycol
PVDF: Polyvinylidene fluoride
PAN: (Polyacrylonitrile)
PVA-g-PAM: Poly (vinyl alcohol)-g-poly (acrylamide)
PAA: Poly (acrylic acid)
CA: Cellulose acetate
CS: Chitosan
SF: Silk fibroin
CSilkNM: Carbonized silk nanofiber membranes
BC: Bacterial cellulose
PHMB-BC: Polyhexamethylene biguanidine bacterial cellulose
St: Starch
Gl: Glycerol
3D: Three-dimensional
2D: Two-dimensional
PSBMA: Poly (sulfobetaine methacrylate)
O-HES: Oxidized hydroxyethyl starch
M-CMCS: Modified carboxymethyl chitosan
TATAPA: Tannic acid, Thioctic acid and Phytic acid
HLTMs: High light-transmitting fibrous membranes

## Acknowledgment

This work was undertaken, in part, thanks to funding from the Canada Research Chairs Program. The authors are grateful for technical support from nanoFAB, at the University of Alberta.

## Conflicts of interest

There are no conflicts to declare.

## Notes and references

1  E. A. Kamoun, E.-R. S. Kenawy and X. Chen, *Journal of advanced research*, 2017, **8**, 217–233.
2  R. G. Sibbald, H. L. Orsted, P. M. Coutts and D. H. Keast, *Advances in skin & wound care*, 2007, **20**, 390–405.
3  M. Abrigo, S. L. McArthur and P. Kingshott, *Macromolecular bioscience*, 2014, **14**, 772–792.
4  D. Simões, S. P. Miguel, M. P. Ribeiro, P. Coutinho, A. G. Mendonça and I. J. Correia, *European Journal of Pharmaceutics and Biopharmaceutics*, 2018, **127**, 130–141.
5  C. Yu, Z.-Q. Hu and R.-Y. Peng, *Military Medical Research*, 2014, **1**, 1–8.
6  C. R. Cardoso, S. Favoreto Jr, L. L. d. Oliveira, J. O. Vancim, G. B. Barban, D. B. Ferraz and J. S. d. Silva, *Immunobiology*, 2011, **216**, 409–415.
7  E. Rezvani Ghomi, S. Khalili, S. Nouri Khorasani, R. Esmaeely Neisiany and S. Ramakrishna, *Journal of Applied Polymer Science*, 2019, **136**, 47738.
8  J. P. Junker, R. A. Kamel, E. Caterson and E. Eriksson, *Advances in wound care*, 2013, **2**, 348–356.
9  D. R. Seshadri, N. D. Bianco, A. N. Radwan, C. A. Zorman and K. M. Bogie, *IEEE Journal of Translational Engineering in Health and Medicine*, 2022, **10**, 1–9.
10  R. Xu, H. Xia, W. He, Z. Li, J. Zhao, B. Liu, Y. Wang, Q. Lei, Y. Kong, Y. Bai et al., *Scientific reports*, 2016, **6**, 1–12.
11  P. Mistry, R. Chhabra, S. Muke, A. Narvekar, S. Sathaye, R. Jain and P. Dan-dekar, *Materials Science and Engineering: C*, 2021, **119**, 111316.
12  M. Varsei, N. R. Tanha, M. Gorji and S. Mazinani, *Polymers and Polymer Com-posites*, 2021, **29**, S1403–S1413.
13  C. Liu, P.-C. Hsu, H.-W. Lee, M. Ye, G. Zheng, N. Liu, W. Li and Y. Cui, *Nature communications*, 2015, **6**, 6205.
14  S. F. Bernatchez, *Journal of the Association for Vascular Access*, 2014, **19**, 256–261.
15  S. P. Kumar, Y. Asokan, K. Balamurugan and B. Harsha, *Journal of Medical Engineering & Technology*, 2022, **46**, 318–334.
16  S. a. Guo and L. A. DiPietro, *Journal of dental research*, 2010, **89**, 219–229.
17  V. Brumberg, T. Astrelina, T. Malivanova and A. Samoilov, *Biomedicines*, 2021, **9**, 1235.
18  N. Eslahi, A. Mahmoodi, N. Mahmoudi, N. Zandi and A. Simchi, *Polymer Re-views*, 2020, **60**, 144–170.

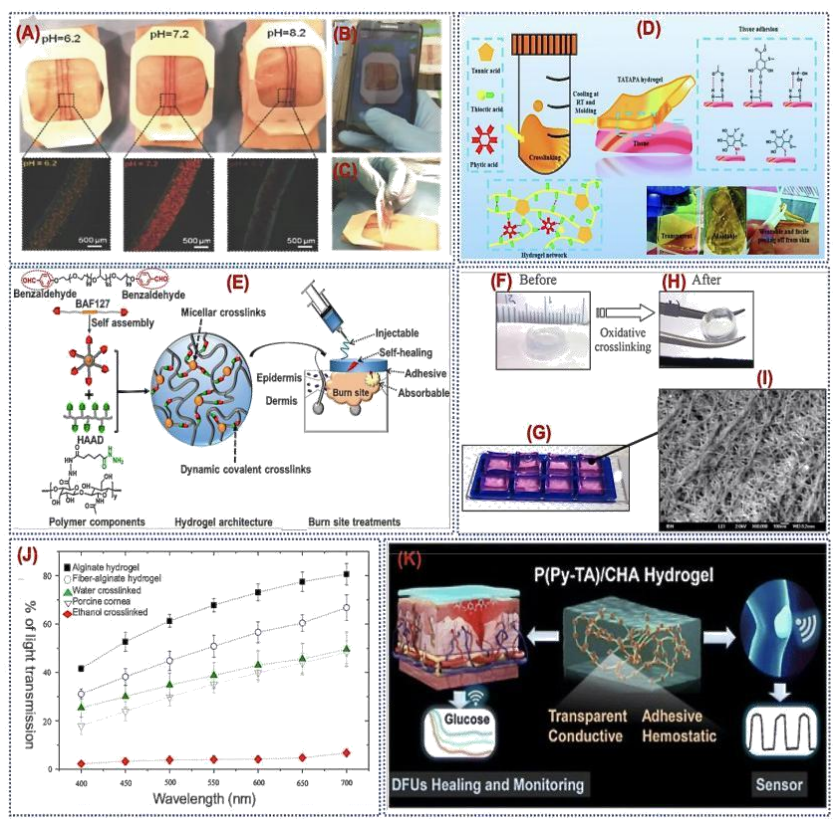

Fig. 10 Materials, dressing, and their characteristics of transparent wound dressing Fabrication of a pH-sensitive wound dressing and its characterization; (A) Fabricated wound dressings are placed on pieces of pig skin sprayed with solutions of different pH, The images confirm sufficient visual difference for identifying the variation in the skin pH for the range relevant to the values in chronic wounds, The insets are showing the images of the engineered fibers in the corresponding solutions, (B) Pictures were taken using a smartphone for determining the pH of the substrate, (C) Flexibility of the fabricated wound dressing and capability for forming conformal contact with skin [190]. Adapted with permission from ref [190], Copyright 2016, Wiley-VCH Verlag GmbH & Co. KGaA, Weinheim, Germany.) (D) Preparation of the TATAPA hydrogel, its appearance, macromolecular network, and tissue adhesion mechanism [191]. (Adapted with permission from ref [191], Copyright 2022, Royal Society of Chemistry, London, United Kingdom.) (E) Schematic presentation of double crosslinked HA-az-F127 hydrogel preparations and burn site treatments involving multiple biological and physical functions [192]. (Adapted with permission from ref [192], Copyright 2018, ACS, New York, United States.) (F-G) After crosslinking, the gel structural rigidity improved and it is now light-transparent and easily pickable with a pair of forceps, (H) LK6C + CRGD gels (99 % water) were translucent, clear, and manageable (phenol red dye added to improved visualization), (I) A FESEM image of gel fibrous network [193]. (Adapted with permission from ref [193], Copyright 2016, Nature Portfolio, Berlin, Germany.) (J) Percentage of light transmission through the porcine cornea and alginate hydrogels that contain different types of randomly-oriented fibers [194]. (Adapted with permission from ref [194], Copyright 2013, Elsevier, Amsterdam, Netherlands.)

(K) A smart hydrogel for all-around treatment and glucose monitoring of diabetic foot ulcers. (Adapted with permission from ref [195], Copyright 2022, Royal Society of Chemistry, London, United Kingdom.)

19  M. Clark, *Alginates and Their Biomedical Applications*, 2018, 213–222.
20  A. Gaspar-Pintiliescu, A.-M. Stanciuc and O. Craciunescu, *International journal of biological macromolecules*, 2019, **138**, 854–865.
21  R. Dong and B. Guo, *Nano Today*, 2021, **41**, 101290.
22  L. Zhang, M. Liu, Y. Zhang and R. Pei, *Biomacromolecules*, 2020, **21**, 3966–3983.
23  C. Gao, L. Zhang, J. Wang, M. Jin, Q. Tang, Z. Chen, Y. Cheng, R. Yang and G. Zhao, *Journal of Materials Chemistry B*, 2021, **9**, 3106–3130.
24  T. Rousseau, C. Plomion and K. Sandy-Hodgetts, *International Wound Journal*, 2022, **19**, 1456–1462.
25  H. U. Zaman, J. Islam, M. A. Khan and R. A. Khan, *Journal of the mechanical behavior of biomedical materials*, 2011, **4**, 1369–1375.
26  M. Gruppuso, G. Turco, E. Marsich and D. Porrelli, *Applied Materials Today*, 2021, **24**, 101148.
27  P. Montero, M. Flandes-Iparraguirre, S. Musquiz, M. Pérez Araluce, D. Plano, C. Sanmartín, G. Orive, J. J. Gavira, F. Prosper and M. M. Mazo, *Frontiers in Bioengineering and Biotechnology*, 2020, **8**, 955.
28  T. A. Debele and W.-P. Su, *International Journal of Polymeric Materials and Polymeric Biomaterials*, 2022, **71**, 87–108.
29  Z. E. Moore and J. Webster, *Cochrane Database of Systematic Reviews*, 2018.
30  S. Alven, X. Nqoro and B. A. Aderibigbe, *Polymers*, 2020, **12**, 2286.
31  S. Dhivya, V. Padma and E. Santhini, *Wound dressings-A review. BioMedicine (Netherlands), 5 (4), 24–28*, 2015.
32  M. Farahani and A. Shafiee, *Advanced Healthcare Materials*, 2021, **10**, 2100477.
33  M. Ochoa, R. Rahimi, J. Zhou, H. Jiang, C. K. Yoon, D. Maddipatla, B. B. Narakathu, V. Jain, M. M. Oscai, T. J. Morken *et al.*, *Microsystems & nanoengineering*, 2020, **6**, 46.
34  B. Rogina-Car, J. Rogina, E. Govorčin Bajsić and A. Budimir, *Journal of Industrial Textiles*, 2020, **49**, 1100–1119.
35  S. T. Tan, S. Winarto, R. Dosan and P. B. Aisyah, *The Open Dermatology Journal*, 2019, **13**, year.
36  C. Weller and V. Team, *Advanced textiles for wound care*, Elsevier, 2019, pp. 105–134.
37  S. D'Alessandro, A. Magnavacca, F. Perego, M. Fumagalli, E. Sangiovanni, M. Prato, M. Dell'Agli and N. Basilico, *BioMed Research International*, 2019, **2019**, year.
38  C. Dai, S. Shih and A. Khachemoune, *Journal of Dermatological Treatment*, 2020, **31**, 639–648.
39  V. Vivcharenko and A. Przekora, *Applied Sciences*, 2021, **11**, 4114.
40  R. Singla, S. Soni, P. M. Kulurkar, A. Kumari, S. Mahesh, V. Patial, Y. S. Padwad and S. K. Yadav, *Carbohydrate polymers*, 2017, **155**, 152–162.
41  Y. Tang, X. Lan, C. Liang, Z. Zhong, R. Xie, Y. Zhou, X. Miao, H. Wang and W. Wang, *Carbohydrate polymers*, 2019, **219**, 113–120.
42  O. M. Ionescu, A.-T. Iacob, A. Mignon, S. Van Vlierberghe, M. Baican, M. Danu,

43  B. Li, S. Pan, H. Yuan and Y. Zhang, *Composites Part A: Applied Science and Manufacturing*, 2016, **90**, 380–389.
44  B. Liesenfeld, D. Moore, A. Mikhaylova, J. Vella, R. Carr, G. Schultz and G. Ol-derman, Symposium on advanced wound care, 2009.
45  A. Sood, M. S. Granick and N. L. Tomaselli, *Advances in wound care*, 2014, **3**, 511–529.
46  N. Rani Raju, E. Silina, V. Stupin, N. Manturova, S. B. Chidambaram and R. R. Achar, *Pharmaceutics*, 2022, **14**, 1574.
47  A. Malakpour-Permlid, I. Buzzi, C. Hegardt, F. Johansson and S. Oredsson, *Scientific reports*, 2021, **11**, 1–18.
48  Z. Pan, H. Ye and D. Wu, *APL bioengineering*, 2021, **5**, 011504.
49  A. A. Shah, J. Yang, T. Kumar, C. Ayranci and X. Zhang, *Colloids and Surfaces A: Physicochemical and Engineering Aspects*, 2023, 131264.
50  K. Chen, Y. Feng, Y. Zhang, L. Yu, X. Hao, F. Shao, Z. Dou, C. An, Z. Zhuang, Y. Luo *et al.*, *ACS applied materials & interfaces*, 2019, **11**, 36458–36468.
51  Q. Xu, A. Sigen, Y. Gao, L. Guo, J. Creagh-Flynn, D. Zhou, U. Greiser, Y. Dong, F. Wang, H. Tai *et al.*, *Acta biomaterialia*, 2018, **75**, 63–74.
52  J. Wang, C. Zhang, Y. Yang, A. Fan, R. Chi, J. Shi and X. Zhang, *Applied Surface Science*, 2019, **494**, 708–720.
53  C. Echeverría, A. Muñoz-Bonilla, R. Cuervo-Rodríguez, D. López and M. Fernández-García, *ACS Applied Bio Materials*, 2019, **2**, 4714–4719.
54  Y. Zhu, B. Kong, R. Liu and Y. Zhao, *Smart Medicine*, 2022, e20220006.
55  S. P. Miguel, D. R. Figueira, D. Simões, M. P. Ribeiro, P. Coutinho, P. Ferreira and I. J. Correia, *Colloids and surfaces B: Biointerfaces*, 2018, **169**, 60–71.
56  Z. Fu, Y. Zhuang, J. Cui, R. Sheng, H. Tomás, J. Rodrigues, B. Zhao, X. Wang and K. Lin, *Engineered Regeneration*, 2022.
57  A. Wang, M. Shao, F. Yang, C. Shao and C. Chen, *European Polymer Journal*, 2021, **160**, 110803.
58  A. Hou, L. Hu, C. Zheng, K. Xie and A. Gao, *Progress in Organic Coatings*, 2020, **149**, 105940.
59  A. K. Bajpai, A. Vishwakarma and J. Bajpai, *Polymer Bulletin*, 2019, **76**, 3269–3295.
60  N. N. Costa, L. de Faria Lopes, D. F. Ferreira, E. M. L. de Prado, J. A. Severi, J. A. Resende, F. de Paula Careta, M. C. P. Ferreira, L. G. Carreira, S. O. L. de Souza *et al.*, *Materials Science and Engineering: C*, 2020, **109**, 110643.
61  T. Gao, M. Jiang, X. Liu, G. You, W. Wang, Z. Sun, A. Ma and J. Chen, *Polymers*, 2019, **11**, 171.
62  A. Ahsan and M. A. Farooq, *Journal of Drug Delivery Science and Technology*, 2019, **54**, 101308.
63  L. Rathod, S. Bhowmick, P. Patel and K. Sawant, *Journal of Drug Delivery Science and Technology*, 2022, **68**, 103035.
64  H. G. Khajeh, M. Sabzi, S. Ramezani, A. A. Jalili and M. Ghorbani, *Colloids and Surfaces A: Physicochemical and Engineering Aspects*, 2022, **633**, 127891.
65  R. Ravikumar, M. Ganesh, V. Senthil, Y. V. Ramesh, S. L. Jakki and E. Y. Choi, *Journal of Drug Delivery Science and Technology*, 2018, **44**, 342–348.
66  S. P. Miguel, R. S. Sequeira, A. F. Moreira, C. S. Cabral, A. G. Mendonça, P. Fer-reira and I. J. Correia, *European Journal of Pharmaceutics and Biopharmaceu-tics*, 2019, **139**, 1–22.
67  J. Kucinska-Lipka, I. Gubanska, A. Lewandowska, A. Terebieniec, A. Przybytek and H. Cieślinski, *Polymer Bulletin*, 2019, **76**, 2725–2742.
68  S. K. Jaganathan and M. P. Mani, *Journal of Applied Polymer Science*, 2019, **136**, 46942.
69  H. Gholami and H. Yeganeh, *Biomedical Materials*, 2020, **15**, 045001.
70  A. W. Jatoi, *Composites Communications*, 2020, **19**, 103–107.
71  Y. Gao and Q. Ma, *Smart Medicine*, 2022, e20220012.
72  Z. Luo, J. Che, L. Sun, L. Yang, Y. Zu, H. Wang and Y. Zhao, *Engineered Regen-eration*, 2021, **2**, 257–262.
73  D. Gao, Y. Mi and Z. Gao, *Materials Letters*, 2020, **276**, 128237.
74  R. K. Sivamani, B. R. Ma, L. N. Wehrli and E. Maverakis, *Advances in wound care*, 2012, **1**, 213–217.
75  A. Hernández-Rangel and E. S. Martin-Martinez, *Journal of Biomedical Materi-als Research Part A*, 2021, **109**, 1751–1764.
76  Z. Kalaycıoğlu, N. Kahya, V. Adımcılar, H. Kaygusuz, E. Torlak, G. Akın-Evingür and F. B. Erim, *European Polymer Journal*, 2020, **133**, 109777.
77  R. Yu, Y. Yang, J. He, M. Li and B. Guo, *Chemical Engineering Journal*, 2021, **417**, 128278.
78  A. Ullah, S. Ullah, M. Q. Khan, M. Hashmi, P. D. Nam, Y. Kato, Y. Tamada and I. S. Kim, *International journal of biological macromolecules*, 2020, **155**, 479–489.
79  P. Makvandi, G. W. Ali, F. Della Sala, W. I. Abdel-Fattah and A. Borzacchiello, *Carbohydrate Polymers*, 2019, **223**, 115023.
80  Y. Duan, K. Li, H. Wang, T. Wu, Y. Zhao, H. Li, H. Tang and W. Yang, *Carbohy-drate polymers*, 2020, **238**, 116195.
81  N. R. Barros, S. Ahadian, P. Tebon, M. V. C. Rudge, A. M. P. Barbosa and R. D. Herculano, *Materials Science and Engineering: C*, 2021, **119**, 111589.
82  L. Xing, Y. Ma, H. Tan, Y. Guan, S. Li, J. Li, Y. Jia, T. Zhou, X. Niu and X. Hu, *Polymer Testing*, 2019, **79**, 106039.
83  P. Taheri, R. Jahanmardi, M. Koosha and S. Abdi, *International journal of bio-logical macromolecules*, 2020, **154**, 421–432.
84  E. E. Leonhardt, N. Kang, M. A. Hamad, K. L. Wooley and M. Elsabahy, *Nature communications*, 2019, **10**, 1–9.
85  L. Colobatiu, A. Gavan, A.-V. Potarniche, V. Rus, Z. Diaconeasa, A. Mocan, I. Tomuta, S. Mirel and M. Mihaiu, *Reactive and Functional Polymers*, 2019, **145**, 104369.
86  P. Deng, L. Yao, J. Chen, Z. Tang and J. Zhou, *Carbohydrate polymers*, 2022, **276**, 118718.
87  Q. Wei, Y. Wang, H. Wang, L. Qiao, Y. Jiang, G. Ma, W. Zhang and Z. Hu, *Carbohydrate Polymers*, 2022, **278**, 119000.
88  M. Najafiasl, S. Osfouri, R. Azin and S. Zaeri, *Journal of Drug Delivery Science and Technology*, 2020, **57**, 101708.
89  Z. Zheng, J. Qi, L. Hu, D. Ouyang, H. Wang, Q. Sun, L. Lin, L. You and B. Tang, *Biomaterials Advances*, 2022, **134**, 112560.
90  L. Sun, L. Li, Y. Wang, M. Li, S. Xu and C. Zhang, *Journal of Tissue Viability*, 2022, **31**, 180–189.
91  S. Zhu, J. Yu, S. Xiong, Y. Ding, X. Zhou, Y. Hu, W. Chen, Y. Lin and L. Dao, *Journal of Applied Polymer Science*, 2022, **139**, 51623.
92  W. Wahbi, R. Siam, J. Kegere, W. A. El-Mehalmey and W. Mamdouh, *ACS omega*, 2020, **5**, 3006–3015.
93  F. Li, L.-G. Ding, B.-J. Yao, N. Huang, J.-T. Li, Q.-J. Fu and Y.-B. Dong, *Journal of Materials Chemistry A*, 2018, **6**, 11140–11146.
94  H. Zhang, J. Ma, C. Liu, L. Li, C. Xu, Y. Li, Y. Li and H. Tian, *Journal of Haz-ardous Materials*, 2022, **435**, 128965.
95  M. Kuddushi, S. Rajput, A. Shah, J. Mata, V. K. Aswal, O. El Seoud, A. Kumar and N. I. Malek, *ACS applied materials & interfaces*, 2019, **11**, 19572–19583.
96  C. Wang, Y. Luo, X. Liu, Z. Cui, Y. Zheng, Y. Liang, Z. Li, S. Zhu, J. Lei, X. Feng *et al.*, *Bioactive materials*, 2022, **13**, 200–211.
97  X. Ren, C. Yang, L. Zhang, S. Li, S. Shi, R. Wang, X. Zhang, T. Yue, J. Sun and J. Wang, *Nanoscale*, 2019, **11**, 11830–11838.
98  G. Tan, Y. Zhong, L. Yang, Y. Jiang, J. Liu and F. Ren, *Chemical Engineering Journal*, 2020, **390**, 124446.
99  D. Liu, J. Wan, G. Pang and Z. Tang, *Advanced Materials*, 2019, **31**, 1803291.
100 G. Zhang, X. Li, Q. Liao, Y. Liu, K. Xi, W. Huang and X. Jia, *Nature Communi-cations*, 2018, **9**, 2785.
101 S. Chen, J. Lu, T. You and D. Sun, *Coordination Chemistry Reviews*, 2021, **439**, 213929.
102 L.-G. Ding, S. Wang, B.-J. Yao, F. Li, Y.-A. Li, G.-Y. Zhao and Y.-B. Dong, *Ad-vanced Healthcare Materials*, 2021, **10**, 2001821.
103 Y.-Z. Long, X. Yan, X.-X. Wang, J. Zhang and M. Yu, *Electrospinning: Nanofab-rication and applications*, Elsevier, 2019, pp. 21–52.
104 Y. Zhao, Z. Q. Li, L. Yang, H. Liu, R. Yan, L. Xiao, H. Liu, J. Wang, B. Yang *et al.*, *Macromolecular Rapid Communications*, 2020, **41**, 2000441.
105 M. Yu, R.-H. Dong, X. Yan, G.-F. Yu, M.-H. You, X. Ning and Y.-Z. Long, *Macro-molecular Materials and Engineering*, 2017, **302**, 1700002.
106 Y. Zou, P. Wang, A. Zhang, Z. Qin, Y. Li, Y. Xianyu and H. Zhang, *ACS Applied Materials & Interfaces*, 2022, **14**, 8680–8692.
107 B. Ding, X. Wang and J. Yu, *Electrospinning: nanofabrication and applications*, William Andrew, 2018.
108 M. R. Ramezani, Z. Ansari-Asl, E. Hoveizi and A. R. Kiasat, *Fibers and Polymers*, 2020, **21**, 1013–1022.
109 M. Chen, Z. Long, R. Dong, L. Wang, J. Zhang, S. Li, X. Zhao, X. Hou, H. Shao and X. Jiang, *Small*, 2020, **16**, 1906240.
110 A. Karakeçili, B. Topuz, S. Korpayev and M. Erdek, *Materials Science and Engi-neering: C*, 2019, **105**, 110098.
111 A. R. Unnithan, G. Gnanasekaran, Y. Sathishkumar, Y. S. Lee and C. S. Kim, *Carbohydrate polymers*, 2014, **102**, 884–892.
112 S. Jiang, B. C. Ma, J. Reinholz, Q. Li, J. Wang, K. A. Zhang, K. Landfester and D. Crespy, *ACS applied materials & interfaces*, 2016, **8**, 29915–29922.
113 R. Augustine, A. Hasan, V. Yadu Nath, J. Thomas, A. Augustine, N. Kalarikkal, A.-E. A. Moustafa and S. Thomas, *Journal of Materials Science: Materials in Medicine*, 2018, **29**, 1–16.
114 M. Ranjbar-Mohammadi, S. H. Bahrami and M. Joghataei, *Materials Science and Engineering: C*, 2013, **33**, 4935–4943.
115 G. Gundewadi, S. G. Rudra, R. Gogoi, T. Banerjee, S. K. Singh, S. Dhakate and A. Gupta, *Industrial Crops and Products*, 2021, **170**, 113727.
116 B. Lu, T. Li, H. Zhao, X. Li, C. Gao, S. Zhang and E. Xie, *Nanoscale*, 2012, **4**, 2978–2982.
117 R. Ahmed, M. Tariq, I. Ali, R. Asghar, P. N. Khanam, R. Augustine and A. Hasan, *International journal of biological macromolecules*, 2018, **120**, 385–393.
118 J. López-Esparza, L. F. Espinosa-Cristóbal, A. Donohue-Cornejo and S. Y. Reyes-López, *Industrial & Engineering Chemistry Research*, 2016, **55**, 12532–12538.
119 M. Ranjbar-Mohammadi, S. Rabbani, S. H. Bahrami, M. Joghataei and F. Moayer, *Materials Science and Engineering: C*, 2016, **69**, 1183–1191.
120 A. S. Asran, K. Razghandi, N. Aggarwal, G. H. Michler and T. Groth, *Biomacro-molecules*, 2010, **11**, 3413–3421.
121 M. Sadri and S. Arab Sorkhi, *Nanomedicine Research Journal*, 2017, **2**, 100–110.
122 I. H. Ali, I. A. Khalil and I. M. El-Sherbiny, *ACS applied materials & interfaces*, 2016, **8**, 14453–14469.
123 M. Soltanzadeh, S. H. Peighambardoust, B. Ghanbarzadeh, M. Mohammadi and J. M. Lorenzo, *Nanomaterials*, 2021, **11**, 1439.
124 K. Shalumon, K. Anulekha, S. V. Nair, S. Nair, K. Chennazhi and R. Jayakumar,

125 M. Rafiq, T. Hussain, S. Abid, A. Nazir and R. Masood, *Materials Research Express*, 2018, **5**, 035007.
126 D. Han, S. Sherman, S. Filocamo and A. J. Steckl, *Acta biomaterialia*, 2017, **53**, 242–249.
127 J. Nourmohammadi, M. Hadidi, M. H. Nazarpak, M. Mansouri and M. Hasannasab, *Fibers and Polymers*, 2020, **21**, 456–464.
128 H. Kesici Güler, F. Cengiz Çallıoglu̇ and E. Sesli Çetin, *The journal of the Textile Institute*, 2019, **110**, 302–310.
129 Q. Guan, L.-L. Zhou, F.-H. Lv, W.-Y. Li, Y.-A. Li and Y.-B. Dong, *Angewandte Chemie International Edition*, 2020, **59**, 18042–18047.
130 R. Chen, J. Liu, Z. Sun and D. Chen, *Nanofabrication*, 2018, **4**, 17–31.
131 H. S. Sofi, R. Ashraf, A. H. Khan, M. A. Beigh, S. Majeed and F. A. Sheikh, *Materials Science and Engineering: C*, 2019, **94**, 1102–1124.
132 A. Haider, S. Haider and I.-K. Kang, *Arabian Journal of Chemistry*, 2018, **11**, 1165–1188.
133 L. Chen, D. Zhang, K. Cheng, W. Li, Q. Yu and L. Wang, *Journal of Colloid and Interface Science*, 2022, **623**, 21–33.
134 S. Latiyan, T. S. Kumar and M. Doble, *Journal of Tissue Engineering and Regenerative Medicine*, 2022, **16**, 653–664.
135 J. Yin, L. Xu and A. Ahmed, *Advanced Fiber Materials*, 2022, **4**, 832–844.
136 C. Li, X. Luo, L. Li, Y. Cai, X. Kang and P. Li, *International Journal of Biological Macromolecules*, 2022, **209**, 344–355.
137 W.-W. Hu and Y.-T. Lin, *Carbohydrate Polymers*, 2022, **289**, 119440.
138 H. S. Sofi, T. Akram, A. H. Tamboli, A. Majeed, N. Shabir and F. A. Sheikh, *International journal of pharmaceutics*, 2019, **569**, 118590.
139 A. W. Jatoi, H. Ogasawara, I. S. Kim and Q.-Q. Ni, *Materials letters*, 2019, **241**, 168–171.
140 J. I. Kim, J. Y. Kim and C. H. Park, *Scientific reports*, 2018, **8**, 3424.
141 K. Ma, Y. Qiu, Y. Fu and Q.-Q. Ni, *Journal of Materials Science*, 2018, **53**, 10617–10626.
142 Q. Wang, M. Jian, C. Wang and Y. Zhang, *Advanced Functional Materials*, 2017, **27**, 1605657.
143 M. Contardi, J. A. Heredia-Guerrero, G. Perotto, P. Valentini, P. P. Pompa, R. Spanò, L. Goldoni, R. Bertorelli, A. Athanassiou and I. S. Bayer, *European Journal of Pharmaceutical Sciences*, 2017, **104**, 133–144.
144 J. Xia, H. Zhang, F. Yu, Y. Pei and X. S. Luo, *Interfaces*, 2020, **12**, 24370–24379.
145 Y. Xiao, H. Luo, R. Tang and J. Hou, *Polymers*, 2021, **13**, 506.
146 S. Zhang, H. Liu, N. Tang, N. Ali, J. Yu and B. Ding, *Acs Nano*, 2019, **13**, 13501–13512.
147 C. Wang, N. Meng, A. A. Babar, X. Gong, G. Liu, X. Wang, J. Yu and B. Ding, *ACS nano*, 2021, **16**, 119–128.
148 W.-C. Wu, P.-Y. Hsiao and Y.-C. Huang, *Journal of Polymer Research*, 2019, **26**, 1–13.
149 S. Massey, F. Iqbal, A. U. Rehman, M. S. Iqbal and F. Iram, *Journal of Biomaterials Science, Polymer Edition*, 2022, **33**, 481–498.
150 L. Yang, Y. Sun, Q. Zou, T. Lu, W. Wang, M. Ma, Z. He, Q. Liu and C. Ye, *Journal of Biomedical Materials Research Part B: Applied Biomaterials*, 2021, **109**, 1145–1155.
151 Z. Jiang, Y. Wang, L. Li, H. Hu, S. Wang, M. Zou, W. Liu and B. Han, *Macromolecular Bioscience*, 2022, **22**, 2100308.
152 S. Atay and F. Yilmaz Kurt, *The journal of vascular access*, 2021, **22**, 135–140.
153 P. Bainbridge, P. Browning, S. F. Bernatchez, C. Blaser and G. Hitschmann, *The Journal of Vascular Access*, 2021, 11297298211050485.
154 W. Chen, P. Zhang, R. Zang, J. Fan, S. Wang, B. Wang and J. Meng, *Advanced Materials*, 2020, **32**, 1907413.
155 A. Das, S. Bhattacharyya, R. Uppaluri and C. Das, *International journal of biological macromolecules*, 2020, **155**, 260–272.
156 P. P. Patil, J. V. Meshram, R. A. Bohara, S. G. Nanaware and S. H. Pawar, *New Journal of Chemistry*, 2018, **42**, 14620–14629.
157 A. Das, R. Uppaluri and C. Das, *International journal of biological macromolecules*, 2019, **131**, 998–1007.
158 Y. Wang, C. Wang, Y. Xie, Y. Yang, Y. Zheng, H. Meng, W. He and K. Qiao, *Carbohydrate polymers*, 2019, **222**, 114985.
159 Y. Huang, H. Zhao, S. Chen, G. Wan and D. Miao, *Journal of Materials Science*, 2022, 1–16.
160 H. Zhang, X. Luo, H. Tang, M. Zheng and F. Huang, *Materials Science and Engineering: C*, 2017, **79**, 84–92.
161 D. S. K. Gunapu, Y. B. Prasad, V. S. Mudigunda, P. Yasam, A. K. Rengan, R. Korla and S. R. K. Vanjari, *International Journal of Biological Macromolecules*, 2021, **176**, 498–509.
162 J. Kim, S. W. Kim, S. Park, K. T. Lim, H. Seonwoo, Y. Kim, B. H. Hong, Y.-H. Choung and J. H. Chung, *Advanced healthcare materials*, 2013, **2**, 1525–1531.
163 F. Ghiasi and M.-T. Golmakani, *Food Hydrocolloids*, 2023, **135**, 108164.
164 M. M. Delavari and I. Stiharu, *Engineering Proceedings*, 2021, **2**, 31.
165 K. Yasuda, M. Ogushi, A. Nakashima, Y. Nakano and K. Suzuki, *in vivo*, 2018, **32**, 799–805.
166 A. Samadi, S. Azandeh, M. Orazizadeh, V. Bayati, M. Rafienia and M. A. Karami, *Advanced Biomedical Research*, 2021, **10**, year.
167 Y. Liang, J. He and B. Guo, *ACS nano*, 2021, **15**, 12687–12722.
168 Y. Yang, Y. Liang, J. Chen, X. Duan and B. Guo, *Bioactive materials*, 2022, **8**, 341–354.
169 G. Tao, R. Cai, Y. Wang, H. Zuo and H. He, *Materials Science and Engineering: C*, 2021, **119**, 111597.
170 H. Jeong, D. Y. Lee, D. H. Yang and Y.-S. Song, *Macromolecular Research*, 2022, **30**, 223–229.
171 Y. Ren, S. Ma, D. Zhang, S. Guo, R. Chang, Y. He, M. Yao and F. Guan, *International Journal of Biological Macromolecules*, 2022, **210**, 218–232.
172 X. Huang, C. Ma, Y. Xu, J. Cao, J. Li, J. Li, S. Q. Shi and Q. Gao, *Industrial Crops and Products*, 2022, **182**, 114945.
173 S. Tavakoli and A. Klar, *Advanced hydrogels as wound dressings. Biomolecules.(2020); 10 (8): 1169*.
174 S. A. Shah, M. Sohail, S. Khan, M. U. Minhas, M. De Matas, V. Sikstone, Z. Hussain, M. Abbasi and M. Kousar, *International journal of biological macromolecules*, 2019, **139**, 975–993.
175 D. K. Pandey, M. Kuddushi, A. Kumar and D. K. Singh, *Colloids and Surfaces A: Physicochemical and Engineering Aspects*, 2022, **650**, 129631.
176 M. Kuddushi, D. K. Pandey, D. K. Singh, J. Mata and N. Malek, *Materials Advances*, 2022, **3**, 632–646.
177 M. Kuddushi, D. Ray, V. Aswal, C. Hoskins and N. Malek, *ACS Applied Bio Materials*, 2020, **3**, 4883–4894.
178 M. Kuddushi, N. K. Patel, S. Rajput, O. A. El Seoud, J. P. Mata and N. I. Malek, *ChemSystemsChem*, 2020, **2**, e1900053.
179 A. R. Abbasi, M. Sohail, M. U. Minhas, T. Khaliq, M. Kousar, S. Khan, Z. Hussain and A. Munir, *International journal of biological macromolecules*, 2020, **155**, 751–765.
180 G. Chen, Y. Yu, X. Wu, G. Wang, J. Ren and Y. Zhao, *Advanced Functional Materials*, 2018, **28**, 1801386.
181 M. Li, Y. Liang, J. He, H. Zhang and B. Guo, *Chemistry of Materials*, 2020, **32**, 9937–9953.
182 G. Xie, N. Zhou, S. Du, Y. Gao, H. Suo, J. Yang, J. Tao, J. Zhu and L. Zhang, *Fundamental Research*, 2022, **2**, 268–275.
183 G. Xie, N. Zhou, Y. Gao, S. Du, H. Du, J. Tao, L. Zhang and J. Zhu, *Chemical Engineering Journal*, 2021, **403**, 126353.
184 H. Chen, R. Cheng, X. Zhao, Y. Zhang, A. Tam, Y. Yan, H. Shen, Y. S. Zhang, J. Qi, Y. Feng *et al.*, *NPG Asia Materials*, 2019, **11**, 3.
185 H. He, Z. Xiao, Y. Zhou, A. Chen, X. Xuan, Y. Li, X. Guo, J. Zheng, J. Xiao and J. Wu, *Journal of Materials Chemistry B*, 2019, **7**, 1697–1707.
186 Y. Zhang, H. Chen, Y. Li, A. Fang, T. Wu, C. Shen, Y. Zhao and G. Zhang, *Polymer Testing*, 2020, **87**, 106517.
187 N. Thet, D. Alves, J. Bean, S. Booth, J. Nzakizwanayo, A. Young, B. V. Jones and A. T. A. Jenkins, *ACS applied materials & interfaces*, 2016, **8**, 14909–14919.
188 J. Li, F. Yu, G. Chen, J. Liu, X.-L. Li, B. Cheng, X.-M. Mo, C. Chen and J.-F. Pan, *ACS applied materials & interfaces*, 2020, **12**, 2023–2038.
189 T. T. Reddy, A. Kano, A. Maruyama, M. Hadano and A. Takahara, *Biomacromolecules*, 2008, **9**, 1313–1321.
190 A. Tamayol, M. Akbari, Y. Zilberman, M. Comotto, E. Lesha, L. Serex, S. Bagherifard, Y. Chen, G. Fu, S. K. Ameri *et al.*, *Advanced healthcare materials*, 2016, **5**, 711–719.
191 X.-h. Shao, X. Yang, Y. Zhou, Q.-c. Xia, Y.-p. Lu, X. Yan, C. Chen, T.-t. Zheng, L.-l. Zhang, Y.-n. Ma *et al.*, *Soft Matter*, 2022, **18**, 2814–2828.
192 Z. Li, F. Zhou, Z. Li, S. Lin, L. Chen, L. Liu and Y. Chen, *ACS applied materials & interfaces*, 2018, **10**, 25194–25202.
193 W. Y. Seow, G. Salgado, E. B. Lane and C. A. Hauser, *Scientific Reports*, 2016, **6**, 1–12.
194 K. Tonsomboon and M. L. Oyen, *Journal of the mechanical behavior of biomedical materials*, 2013, **21**, 185–194.
195 H. Liu, Z. Li, S. Che, Y. Feng, L. Guan, X. Yang, Y. Zhao, J. Wang, A. V. Zvyagin, B. Yang *et al.*, *Journal of Materials Chemistry B*, 2022, **10**, 5804–5817.

Table 1 The synthetic polymers for transparent wound dressings

| Polymer | Structure | Types of dressing | Functionality | Main findings | Ref. |
|---|---|---|---|---|---|
| PVDF | 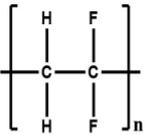 | Composite film | Wound infection control | Nanoparticles addition improved anti bacterialactivity. | 57 |
| PU | 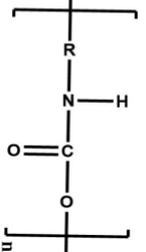 | Composite film | Good antibacterial activity | Composite shows good thermal, hydrophilicity, & mechanical properties. | 58 |
| | | Electrospun composite | Good mechanical strength | Optically transparent nanofiber and %T for post treated nanofiber as 84.0% | 49 |
| PVA-g-PAM | 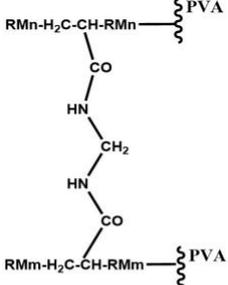 | Hydrogel | Good swelling | Hydrogel shows antibacterial nature against gram-positive bacteria. | 59 |
| PVA-Starch-PAA | 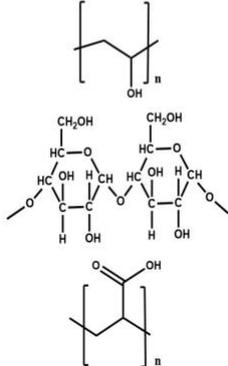 | Film | Antimicrobial and healing activities | These films help complete wound closure | 60 |
| PVA | (structure) | Hydrogel | Adhesion and proliferation feature | Promote wound healing, but also the safe application of stem cells. | 61 |
| | | Hydrogel patches | Good antibacterial activity | These hydrogel patches exhibit excellent wound-healing activity | 62 |
| | | Hydrogel | Stable physicochemical properties | Convenient and effective wound dressing with an ideal feature | 63 |
| | | Hydrogel | Superb antibacterial & biocompatibility | Great potential for the treatment of bacterial infections and significantly accelerated wound healing | 52 |
| PEG | (structure) | Hydrogel | Stable mechanical, non swelling, antifouling properties | A new injectable hydrogel platform as a stem cell delivery and retention system | 51 |
| | | Hydrogel | Highly stretchable, adhesive, and self-healing | Great potential for the treatment of bacterial infections and significantly accelerated wound healing | 50 |
| PLA | (structure) | Electrospun composite | Good antibacterial properties | A new injectable hydrogel platform as a stem cell delivery and retention system. | 53 |
| PU/PAA | (structure) | Electrospun composite | Good mechanical, swelling ratio, and proliferation | The drug thiazolium helps bacteria kill by contact. | 64 |
| PCL/PEG | (structure) | Electrospun nanofiber patch | Uniform fiber and size distribution | Prepared nanofibers accelerate the wound-healing process. | 65 |

Table 2 The natural polymers used to prepare transparent wound dressings

| Polymer | Structure | Types of dressing | Functionality | Main findings | Ref. |
|---|---|---|---|---|---|
| Chitosan | 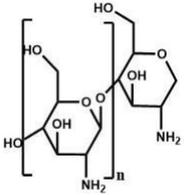 | Films | Antioxidant activity, proliferative effect, and good biocompatibility | Chitosan film dressing the has potential to accelerate wound healing. | 85 |
| | | Hydrogel | Adenine-modified | Adenine-modified chitosan hydrogels reduce inflammatory cell infiltration & accelerate wound healing | 86 |
| | | Hydrogel | Antibacterial and antioxidant properties | The hydrogel exhibits cyto-compatibility. | 87 |
| Alginate | 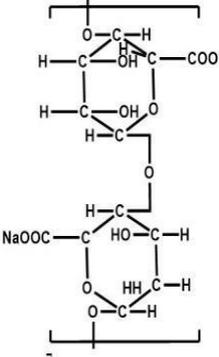 | Electrospun nanofibers | Good biocompatibility and swelling tendency | Promising for wound healing and can enhance specific tissues' regeneration. | 88 |
| | | Hydrogel | A potential | A potential candidate for skin tissue engineering with antioxidant and anti inflammatory effect | 89 |
| | | Casting membrane | Improved tensile strength, swelling rate, water vapor permeability | These dressings have a promising future in the treatment of bacterial infections as wound dressing. | 82 |
| Silk fibroin | 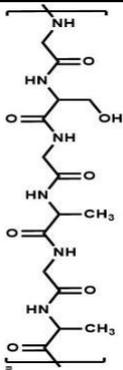 | Hydrogel | Good self-healing, injectability, and biocompatibility. | The hydrogel showed improved wound healing efficacy In vivo model. | 77 |
| Collagen | 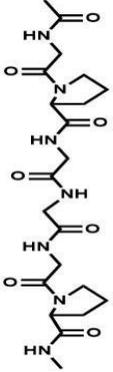 | Bi-layered composite | Enhance cell viability and proliferation | Stimulated wound healing & accelerated re-epithelialization | 90 |
| | | Hydrogel | Antimicrobial activity | Enhanced antibacterial and biocompatibility via arginine and dopamine modification. | 91 |
| Natural rubber latex | 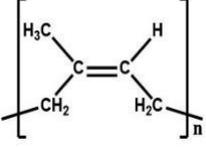 | Membrane | Cell adhesion, and accelerate wound healing | NRL-alginate dressings have great potential to improve diabetic wound care | 81 |
| Fibrinogen | - | Electrospun scaffolds | Good biocompatibility and %T for Chit/CelAc and Chit/CelAc/CeO2-0.1 as 31.52% and 12.72%, respectively. | Material has great developmental potential in tissue engineering | 76 |
| Inulin | 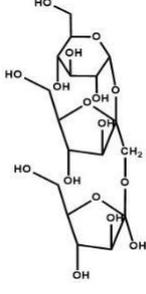 | Electrospun nanofibers | Excellent Prebiotic and Antibacterial | Promising biomaterial for wound healing with antibacterial properties | 92 |

Table 3 Antibacterial polymeric electrospun nanofibers as a wound dressing application

| Polymer | Structure | Antibacterial agent | Model Used In vivo / In vitro | Bacterial Microorganism | Ref. |
|---|---|---|---|---|---|
| PU/CA | 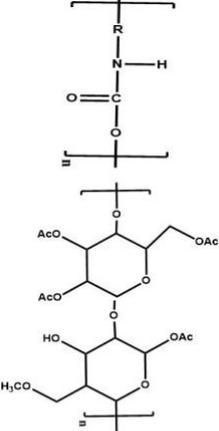 | Zein +Antibiotic drug | In vitro | E.coli:12mm; B.subtilis:15mm; S. aureus: 8mm | 111 |
| PVA | 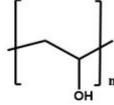 | Octyl methoxy cinnamate (OMC), peppermint oil, amphiphilic octenidine | In vitro | 99% resistance against E. coli K-12 and B. Subtilis | 112 |
| | | Ag NPs | In vitro | Good inhibition zone against E. coli | 113 |
| | | Gum tragacanth | In vitro | Capability to resist P. aeruginosa and S. aureus | 114 |
| | | 1:1 blend of thyme and betel leaf essential oil | In vitro | Capability to inhibit the growth of inoculated C.gloeosporioides and reduce the disease incidence from 100 % in case of control to 40 %. | 115 |
| PVA/CS | 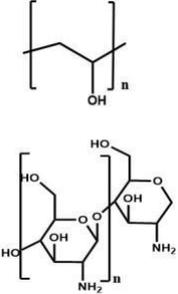 | Graphene | In vitro | Resistance against E. coli | 116 |
| | | ZnO | In vitro In vivo | Higher inhibition zone against E. coli, P. aeroginosa, B. subtilis, and S. aureus compared with pristine CS/PVA NFs | 117 |
| PCL | 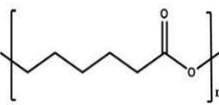 | Ag NPs | In vitro | Adequate resistance to bacteria like S. aureus, E. coli, P. aeruginosa, S. pyogenes, and K. pneumonia | 118 |
| | | Curcumin and gum tragacanth | In vitro In vivo | Antibacterial activity of 99.99% and 85.14% against GNB (MRSA) and GPB (extended-spectrum b lactamase–ESBL) | 119,120 |
| CS / PEO | 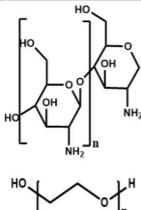 | Cefazolin | In vitro, In vivo | S.aureus: 12mm E. coli : 10mm | 121 |
| | | Ag NPs | In vitro | Higher inhibition zone than pristine CS/PEO NF (0.01mm) | 122 |
| | | pomegranate peel | In vitro | Resistance against E. coli, S. aureus and S. epidermidis | 60,123 |
| Sodium Alginate / PVA | 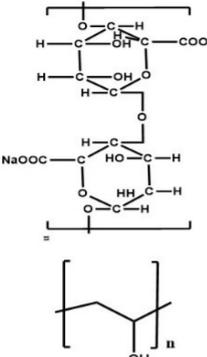 | ZnO | In vitro | S. aureus: 15-16mm E. coli bacteria: 14–15mm | 124 |
| | | Essential oils (cinnamon, clove, and lavender) | In vitro | High antibacterial properties against S. aureus | 125 |
| CA/ PCL/PVP | 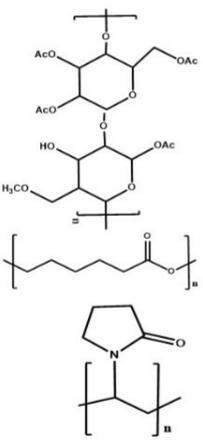 | Nisin | In vitro | High antimicrobial activity | 126 |
| SF-PVA | 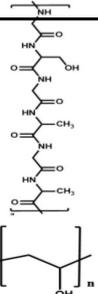 | Elaeagnus Angustifolia (EA) | In vitro | Antibacterial activity against both GPB (S. aureus) and GNB (E. coli) | 127 |
| PVP | 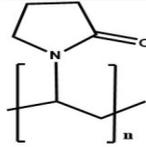 | Cinnamon oil | In vitro | Good inhibition zone against S.aureus, E.coli, C.albicans, and P.auruginosa | 128 |

Table 4 Electrospinning experimental parameters for fabrication of transparent nanofibers

| Materials | Experimental Parameters | Properties of nanofiber | Ref. |
|---|---|---|---|
| PCL/collagen | Applied voltage = 15 kV, distance from the needle tip to the substrate = 15 cm, and solution flow rate = 1 ml/hr. | High transparency and with hemispherical design | 140 |
| PCL/shellac | PCL solution: Voltage = 12 kV, distance from the needle tip to the substrate = 15 cm, and solution flow rate = 0.3 mm/min. shellac solution: Voltage = 15 kV, distance from the needle tip to the substrate = 20 cm, and solution flow rate = 0.5 mm/min. | Good mechanical properties and mensurable transparency. | 141 |
| Silk | Applied voltage = 20 kV, distance from the needle tip to the substrate = 20 cm, and solution flow rate = 1 ml/hr. | Low-cost and large-scale capable approach, Flexible, transparent, with ultrathin structures | 142 |
| PU | Applied voltage = 11 kV, distance from the needle tip to the substrate = 15 cm, and solution flow rate = 1.5 ml/hr. | Transparent, hydrophobic, and mechanically robust composite nanofibrous membranes | 49 |
| polyvinylpyrrolidone (PVP) | Applied voltage = 30.5 kV, distance from the needle tip to the substrate = 24 cm, and solution flow rate = 250 ml/hr. | Antibiotic activity, high transparency, stretchable, and soft. | 143 |
| CS/cellulose | Applied voltage = 200 kV, distance from the needle tip to the substrate = 10 cm, solution flow rate = 2 ml/hr, and needle inner diameter of 0.6 mm. | Transparent, porous, self-healable | 144 |